\documentclass[aps,showpacs,preprint,superscriptaddress]{revtex4}
\usepackage{graphicx}
\usepackage{subfigure}
\usepackage{float}
\usepackage{amsmath}
\usepackage{amsfonts}
\usepackage{bm}
\usepackage{bbm}
\usepackage{txfonts}
\usepackage{array}
\usepackage{amssymb}

\begin{document}

\title{Pair production in differently polarized electric fields with frequency chirps}
\author{Obulkasim Olugh}
\affiliation{Key Laboratory of Beam Technology of the Ministry of Education, and College of Nuclear Science and Technology, Beijing Normal University, Beijing 100875, China}
\author{Zi-Liang Li}
\affiliation{School of Science, China University of Mining and Technology, Beijing 100083, China}
\author{Bai-Song Xie \footnote{bsxie@bnu.edu.cn}}
\affiliation{Key Laboratory of Beam Technology of the Ministry of Education, and College of Nuclear Science and Technology, Beijing Normal University, Beijing 100875, China}
\affiliation{Beijing Radiation Center, Beijing 100875, China}
\author{Reinhard Alkofer \footnote{reinhard.alkofer@uni-graz.at}}
\affiliation{Institute of Physics, University of Graz, NAWI Graz, Universit\"atsplatz 5, 8010 Graz, Austria}

\date{\today}
\begin{abstract}
Electron-positron pair production in strong electric fields, {\it i.e.}, the Sauter-Schwinger
effect, is studied using the real-time Dirac-Heisenberg-Wigner formalism. Hereby, the electric
field is modeled to be a homogeneous, single-pulse field with subcritical peak field strength.
Momentum spectra are calculated for four different polarizations  - linear, elliptic,
near-circular elliptic or  circular - as well as a number of linear frequency chirps.
With details depending on the chosen polarization the frequency chirps lead to strong
interference effects  and thus quite  substantial changes in the momentum spectra.
The resulting produced  pairs' number densities depend non-linearly on the parameter characterizing the polarization and are very sensitive to variations of the chirp parameter. For some of the investigated frequency chirps this can
provide an enhancement of the number density by three to four orders of magnitude.
\end{abstract}
\pacs{12.20.Ds, 03.65.Pm, 02.60.-x}
\maketitle

\section{Introduction}

Electron-positron ($e^{+}e^{-}$) pair production in strong electric fields, also
known as the Sauter-Schwinger effect, is a long-standing theoretical prediction
\cite{Sauter:1931zz,Heisenberg:1935qt,Schwinger:1951nm}, which is,
however, not yet experimentally verified, for a recent review see, {\it e.g.},
\cite{Gelis:2015kya}. The pair production rate is hereby exponentially suppressed and
proportional to $\exp (-\pi E_{cr} /E )$ as long as the electric field is of the order
of or smaller than the critical field,
$E_{cr} =  {m_e^2c^3} / {e\hbar} \approx 1.3 \cdot 10^{18}  {\rm V}/{\rm m}$.
The related laser intensity, {\it e.g.},  $I=4.3 \times10^{29}$W/cm$^{2}$ for $1\mu$m
light, is beyond current technological possibilities but the progress in
high-intensity laser technology \cite{Heinzl:2008an,Marklund:2008gj,Pike:2014wha}
might make experimental tests possible in the next decade, especially in view of
planned facilities as the Extreme Light Infrastructure (ELI), the Exawatt Center
for Extreme Light Studies (XCELS), or the Station of Extreme Light at the Shanghai
Coherent Light Source. On the other hand, the already operating
X-ray free electron laser (XFEL)  at DESY in Hamburg can in principle achieve
near-critical field strength as large as $E\approx 0.1\, E_{cr}$, see,
{\it e.g.},  \cite{Ringwald:2001ib}. Triggered by the technical design report
of the XFEL numerical estimates of the achievable number densities and of the
resulting momentum spectra have been performed within the quantum kinetic approach
at the beginning of the millenium \cite{Alkofer:2001ik,Roberts:2002py} but many studies
of the Sauter-Schwinger effect based on a number of different theoretical techniques
have been undertaken in the last and in this century, for a guide to the literature
we refer to the recent review~\cite{Gelis:2015kya}.

\begin{figure}[hb]
\begin{center}
\includegraphics[width=15cm, height=10cm]{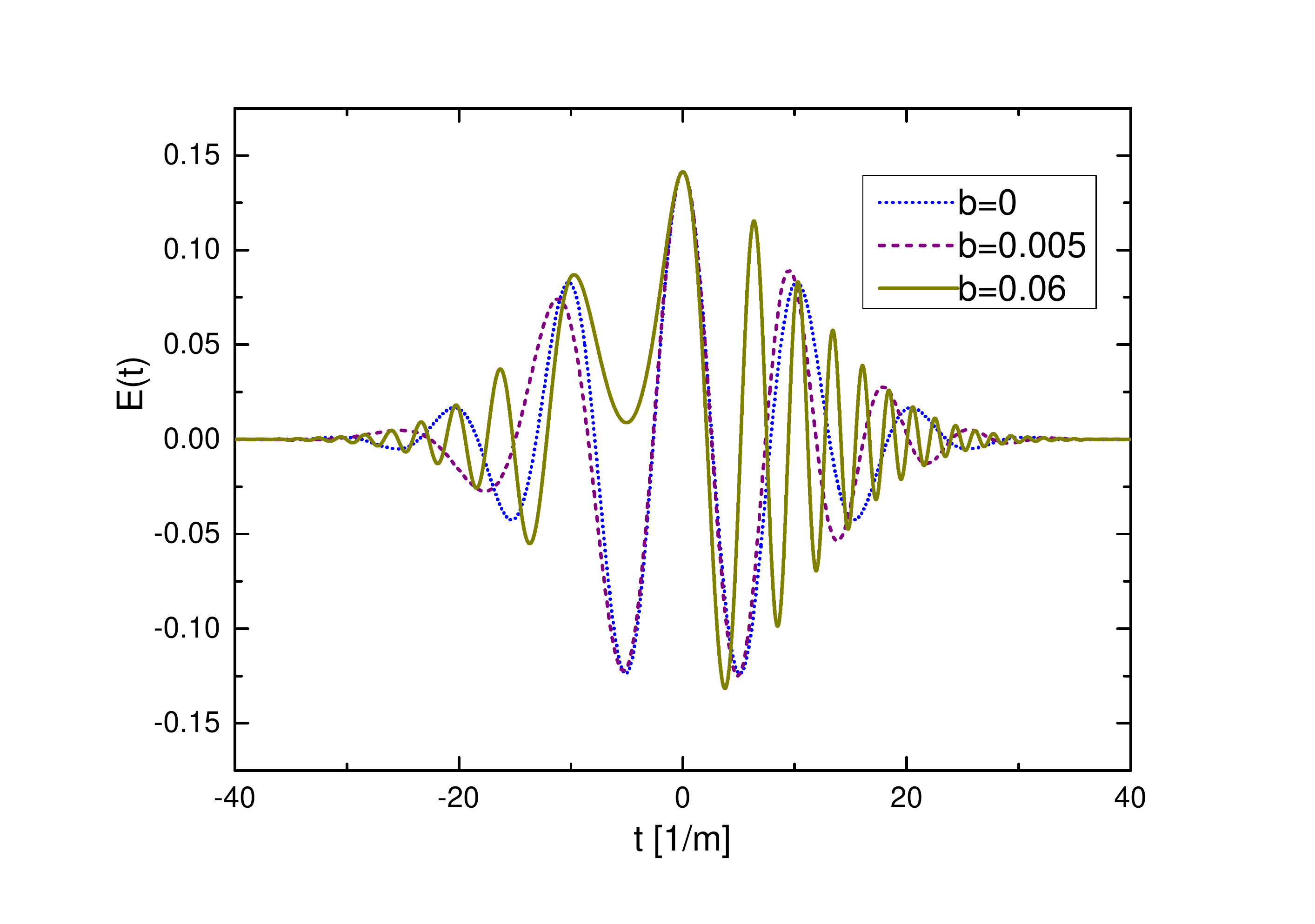}
\end{center}
\caption{The time dependence of the electric field $E(t)$ in units of the critical
field for the linearly polarized ($\delta=0$) case.
The chosen parameters are $E_{0}=0.1\sqrt{2}E_{cr}$, $\omega=0.6m$, and $\tau=10/m$
where $m$ is the electron mass.
The blue dotted line shows the electric field without a chirp, $b=0$.
The purple dashed line displays the field with a chirp parameter $b=0.005$~$m^2$,
the dark yellow-green solid line for the chirp parameter $b=0.06$~$m^2$.}
\label{fig:1}
\end{figure}

Amongst the contributions from theorists towards an experimental verification of
non-perturbative ultra-strong field pair production the dynamically assisted
Sauter-Schwinger effect \cite{Schutzhold:2008pz} deserves special mentioning.
It exploits the idea that a combination of a low with a high frequency laser pulse
leads to $e^{+}e^{-}$ pair production rates which are by several orders
of magnitude larger than the sum of the rates for the two separate pulses.
Herein we report on a study which extends this idea by exploiting time-dependent
frequencies, {\it i.e.}, frequency chirps. We focus on linear chirps but allow
then for different types of polarization. For simplicity we study pair production
in a single-pulse field with a Gaussian envelope:
\begin{equation}\label{eq1}
\mathbf{E}(t) \,\, =\, \, \frac{E_{0}}{\sqrt{1+\delta^{2}}}\,
\exp\left(-\frac{t^2}{2\tau^2}\right) \,
\left(\begin{array}{c}
          \cos(b t^{2}+\omega t+\phi) \\
           \delta\sin(b t^{2}+\omega t+\phi) \\
              0 \\
        \end{array}\right),
\end{equation}
where ${E_{0}}/{\sqrt{1+\delta^{2}}}$ is the amplitude of the electric field,
$\tau$ denotes the pulse duration and $\omega$ the oscillation frequency
at $t=0$. For completeness we kept the carrier phase $\phi$ (which is known to
have a significant effect on the momentum spectra of the produced pair
\cite{Hebenstreit:2009km,Abdukerim:2013vsa}) in this expression, however,
it will be set to zero
in the following. The main interest in this study is the dependence on the chirp
parameter $b$. Note that a non-vanishing $b$ can be interpreted as a time-dependent
effective frequency, $\omega_{\mathrm {eff}} = \omega + b t$. The effect of the chirp
parameter $b$ on the time dependence of the electric field is displayed in
Fig.~\ref{fig:1}. The parameter $\delta$
with $-1\le \delta \le 1$ describes the ellipticity of the electric field, $\delta=0$
corresponds to linear  and $\delta =1$ to circular polarization.

For the here presented
study the Dirac-Heisenberg-Wigner (DHW) formalism adapted to pair production
\cite{Vasak:1987um,Hebenstreit:2011pm} is used. This choice is motivated due to its efficiency for
calculations involving circularly
or elliptically polarized electric background fields, see, {\it e.g.}, \cite{Blinne:2013via,Blinne:2016yzv},
in which pair production in rotating circularly polarized electric fields has been
investigated, or \cite{Li:2015cea} for elliptically polarized fields.

At this point a remark with respect to the chosen polarizations for this study is in
order. On the experimental side, due to limitations in instruments, it is much harder
to produce a perfect circularly polarized field than an elliptically polarized
and/or linearly polarized field. For low-intensity laser fields a polarization of up to
$ \pm 0.93$ have been achieved experimentally \cite{Pfeiffer1}. Due to this we
include calculations for a near-circular elliptic polarization,
{\it i.e.}, for $\delta=0.9$.

In addition, we note that high-intensity laser pulses are also obtained through the chirped laser pulse
amplification technique \cite{Strickland}. Therefore the study of pair
production in fields with frequency chirps is well-motivated even besides the
here found amplification similar to the one of the
dynamically assisted Sauter-Schwinger effect.

Throughout this paper natural units $\hbar=c=1$ are used.
Furthermore, three of the five  parameters characterizing the electric field stay fixed:
\begin{equation}
E_{0} = 0.1 \sqrt{2}\, E_{cr} \, \quad \omega=0.6m \,  \quad \tau=10/m \, ,
\label{FieldParameters}
\end{equation}
 where $m$ is the electron mass.
The Keldysh adiabaticity parameter is thus $\gamma= 4.25 \sqrt{1+\delta^{2}}$.
For the chirp parameter $b$ we investigate several cases in the interval $0 \le b \le 0.06$~$m^2$,
and for the polarization four different values of $\delta$ are chosen.
We are aware that the pulse length is hardly sufficient to provide a clean multi-photon signal,
and that  a value of $b=0.06$~$m^2$ is already to large to be classified as a ``normal
chirp'', however, the goal of the present exploratory study is a qualitative understanding
of the influence of chirps on the produced number densities of pairs and the related
momenta spectra for different polarizations, and to this end the chosen parameter sets
are very suitable.

This paper is organized as follows: In Sec.~\ref{method} we introduce briefly the
DHW formalism to make the presentation reasonably self-contained.
In Sec.~\ref{result1} we present our numerical results for the
number densities for different chirp parameters and different polarizations.
In Sec.~\ref{result2}  we summarize briefly for four polarizations
how generic properties of the momentum spectra
change for an increasing chirp parameter.
In Sec.~\ref{result3} we re-analyze the spectra within a semi-classical treatment and
discuss in how far the momentum spectra can be qualitatively understood.
In the last section we present our conclusions.

\section{Theoretical description: The DHW formalism}\label{method}

The here presented study employs the DHW formalism which is a relativistic
phase-space approach. It has been further developed for the case of Sauter-Schwinger
pair production in refs.~\cite{Vasak:1987um,Hebenstreit:2011pm}.
Within this method the electron is treated as a quantum field but the laser pulse
is approximated by its mean-field which is justified by the magnitude of the
used electric field.

To make this paper self-contained we briefly review the formalism. To this end we
start from the gauge-invariant density operator of the system,
\begin{equation}
 \hat {\mathcal C}_{\alpha \beta} \left( r , s \right) = \mathcal U \left(A,r,s
\right) \ \left[ \bar \psi_\beta \left( r - s/2 \right), \psi_\alpha \left( r +
s/2 \right) \right],
\end{equation}
in terms of the electron's spinor-valued Dirac field $\psi_\alpha (x)$, and
$r$ denotes the center-of-mass  and $s$ the relative coordinate.
The Wilson line factor
\begin{equation}
 \mathcal U \left(A,r,s \right) = \exp \left( \mathrm{i} \ e \ s \int_{-1/2}^{1/2} d
\xi \ A \left(r+ \xi s \right)  \right)
\end{equation}
renders the density operator gauge-invariant. Note that this factor depends on the
elementary charge $e$ and the background gauge field $A$. The background field is
treated in mean-field (Hartree) approximation, {\it i.e.},
\begin{equation}
 F^{\mu \nu} \left( {x} \right) \approx \langle \hat F^{\mu \nu} \left(
{x} \right) \rangle ,
\end{equation}
and, because  in a given Lorentz frame and gauge the background gauge field
$A \left(\mathbf{x}, t \right)$ is a fixed $c$-number valued function,
no path ordering is needed. The covariant Wigner operator,
\begin{equation}
 \hat{\mathcal W}_{\alpha \beta} \left( r , p \right) = \frac{1}{2} \int d^4 s \
\mathrm{e}^{\mathrm{i} ps} \  \hat{\mathcal C}_{\alpha \beta} \left( r , s
\right),
\end{equation}
thus includes the electron's quantum fluctuations but not the one of the electric field.

The simplification introduced by the mean-field approximation for the electromagnetic field
becomes apparent when one considers the vacuum expectation value of the
covariant Wigner operator to obtain the covariant Wigner function
\begin{equation}
 \mathbbm{W} \left( r,p \right) = \langle \Phi \vert \hat{\mathcal W} \left( r,p
\right) \vert \Phi \rangle.
\end{equation}
In the equation of motion of this correlation function the electromagnetic field factors out:
\begin{equation}
 \langle \Phi \vert F_{\mu \nu} \ \hat{\mathcal{C}} \vert \Phi \rangle =
F_{\mu \nu} \langle \Phi \vert \hat{\mathcal{C}}
\vert \Phi \rangle \, .
\end{equation}
This in turn allows to terminate the in general infinite hierarchy of correlation functions.

As the Wigner function is a Dirac-matrix valued quantity it can be decomposed
into 16 covariant Wigner coefficients
\begin{equation}
\mathbbm{W} = \frac{1}{4} \left( \mathbbm{1} \mathbbm{S} + \textrm{i} \gamma_5
\mathbbm{P} + \gamma^{\mu} \mathbbm{V}_{\mu} + \gamma^{\mu} \gamma_5
\mathbbm{A}_{\mu} + \sigma^{\mu \nu} \mathbbm{T}_{\mu \nu} \right) \, .
\label{decomp}
\end{equation}
Hereby, the related spin and parity properties are made evident by the notation.
As the modeling of the electric field already indicates we work in a definite frame.
Correspondingly, one can project on equal times which yields the equal-time Wigner function
\begin{align}
 \mathbbm{w} \left( \mathbf{x}, \mathbf{p}, t \right) = \int \frac{d p_0}{2 \pi}
\ \mathbbm{W} \left( r,p \right)
\end{align}
and by an analogous decomposition to eq. (\ref{decomp})
the corresponding equal-time Wigner coefficients $\mathbbm{s},\mathbbm{p},
\mathbbm{v}_{0,x,y,z}$ etc..

As the equations of motions for the Wigner coeffecients are quite lengthy we refrain
from repeating the respective formula here. Their explicit form as well as detailed derivations
can be found in \cite{Hebenstreit:2011pm,Kohlfurst:2015zxi}.
A decisive advantage of employing Wigner coefficients is given by the relation of
${\mathbbm s}$ with the mass, of ${\mathbbm v}_0$ with the charge,
and of ${\vec {\mathbbm v}}$ as current density in the case without electric field
\cite{Vasak:1987um,Hebenstreit:2011pm}. Correspondingly, one chooses vacuum initial conditions
as starting values. The non-vanishing values are
\begin{equation}
{\mathbbm s}_{vac} = \frac{-2m}{\sqrt{{\mathbf p}^2+m^2}} \, ,
\quad  {\mathbbm v}_{i,vac} = \frac{-2{ p_i} }{\sqrt{{\mathbf p}^2+m^2}} \, .
\end{equation}

In general, the equations of motions for the Wigner coefficients are integro-differential
equations. Their numerical solution is due to the non-local nature of the respective
pseudo-differential operators very challenging, see, {\it e.g.},
\cite{Hebenstreit:2011pm,Kohlfurst:2015zxi,Kohlfurst:2015niu,Berenyi:2017haf}. For the
homogeneous electric field \eqref{eq1} studied here these equations can be reduced to
ordinary differential equations \cite{Blinne:2013via}. To this end we note first that
then at most ten out of the sixteen Wigner coefficients are non-vanishing:
\begin{equation}
{\mathbbm w} = ( {\mathbbm s},{\mathbbm v}_i,{\mathbbm a}_i,{\mathbbm t}_i)
\, , \quad  {\mathbbm t}_i := {\mathbbm t}_{0i} -   {\mathbbm t}_{i0}  \, .
\end{equation}
Second, the kinetic momentum ${\mathbf p} $ is related to the canonical momentum
${\mathbf q}$ via
\begin{equation}
{\mathbf p}(t) = {\mathbf q} - e {\mathbf A} (t)
\end{equation}
and thus time-dependent. In a next step one expresses the scalar Wigner coefficient
by the one-particle distribution function $f({\mathbf q},t)$. The latter is related
to the phase-space energy density,
\begin{equation}
\varepsilon = m {\mathbbm s} + p_i {\mathbbm v}_i \, .
\end{equation}
via
\begin{equation}
f({\mathbf q},t) = \frac 1 {2 \Omega(\mathbf{q},t)} (\varepsilon - \varepsilon_{vac} ).
\end{equation}
Hereby, $\Omega(\mathbf{q},t)= \sqrt{{\mathbf p}^2(t)+m^2}=
\sqrt{m^{2}+(\mathbf{q}-e\mathbf{A}(t))^{2}}$ is the electron's (resp., positron's)
energy.

In addition, it is helpful to define an auxiliary three-dimensional vector
$\mathbf{v}(\mathbf{q},t)$:
\begin{equation}
v_i (\mathbf{q},t) : = {\mathbbm v}_i (\mathbf{p}(t),t) -
(1-f({\mathbf q},t))  {\mathbbm v}_{i,vac} (\mathbf{p}(t),t) \, .
\end{equation}
The one-particle
momentum distribution function $f(\mathbf{q},t)$ can be then be obtained by solving
the following ten ordinary differential equations for $f(\mathbf{q},t)$ and the nine auxiliary
quantities $v_i(\mathbf{q},t),
a_i(\mathbf{q},t):=\mathbbm{a}_i(\mathbf{q},t)$ and
$t_i(\mathbf{q},t):=\mathbbm{t}_i(\mathbf{q},t)$:
\begin{equation}
\begin{array}{l}
\displaystyle
\dot{f}=\frac{e}{2\Omega} \, \,  \mathbf{E}\cdot \mathbf{v},\\[2mm]
\displaystyle
\dot{\mathbf{v}}=\frac{2}{\Omega^{3}}
\left( (e\mathbf{E}\cdot \mathbf{p})\mathbf{p}-e\Omega^{2}\mathbf{E}\right) (f-1)
-\frac{(e\mathbf{E}\cdot \mathbf{v})\mathbf{p}}{\Omega^{2}}
-2\mathbf{p}\times \mathbf{a} -2m \mathbf{t},\\[2mm]
\displaystyle
\dot{\mathbf{a}}=-2\mathbf{p}\times \mathbf{v},\\
\displaystyle
\dot{\mathbf{t}}=\frac{2}{m}[m^{2}\mathbf{v}-(\mathbf{p}\cdot \mathbf{v})\mathbf{p}],
\end{array}
\end{equation}
where as usual the dot is a shorthand for the time derivative.
Together with the initial conditions
$f(\mathbf{q},-\infty)=0$,
$\mathbf{v}(\mathbf{q},-\infty)=
\mathbf{a}(\mathbf{q},-\infty)=\mathbf{t}(\mathbf{q},-\infty)=0$,
this set of equation is a well-defined and numerically straightforward solvable initial
value problem.

The number density of created pairs is obtained by integrating the distribution function
$f(\mathbf{q},t)$ over all momenta at asymptotically late times $t\to +\infty$:
\begin{equation}\label{3}
  n = \lim_{t\to +\infty}\int\frac{d^{3}q}{(2\pi)^ 3}f(\mathbf{q},t) \, .
\end{equation}

\section{Numerical results for number densities}\label{result1}

As already stated the carrier phase is chosen to be $\phi=0$ leaving
studies similar to the one presented here but
with non-vanishing carrier phase for future investigations.
Herein, we examine the main results for the number density of the
produced particles for several chirp parameters for the different polarization.

\begin{figure}[ht]
\includegraphics[width=\textwidth,height=0.75\textheight]{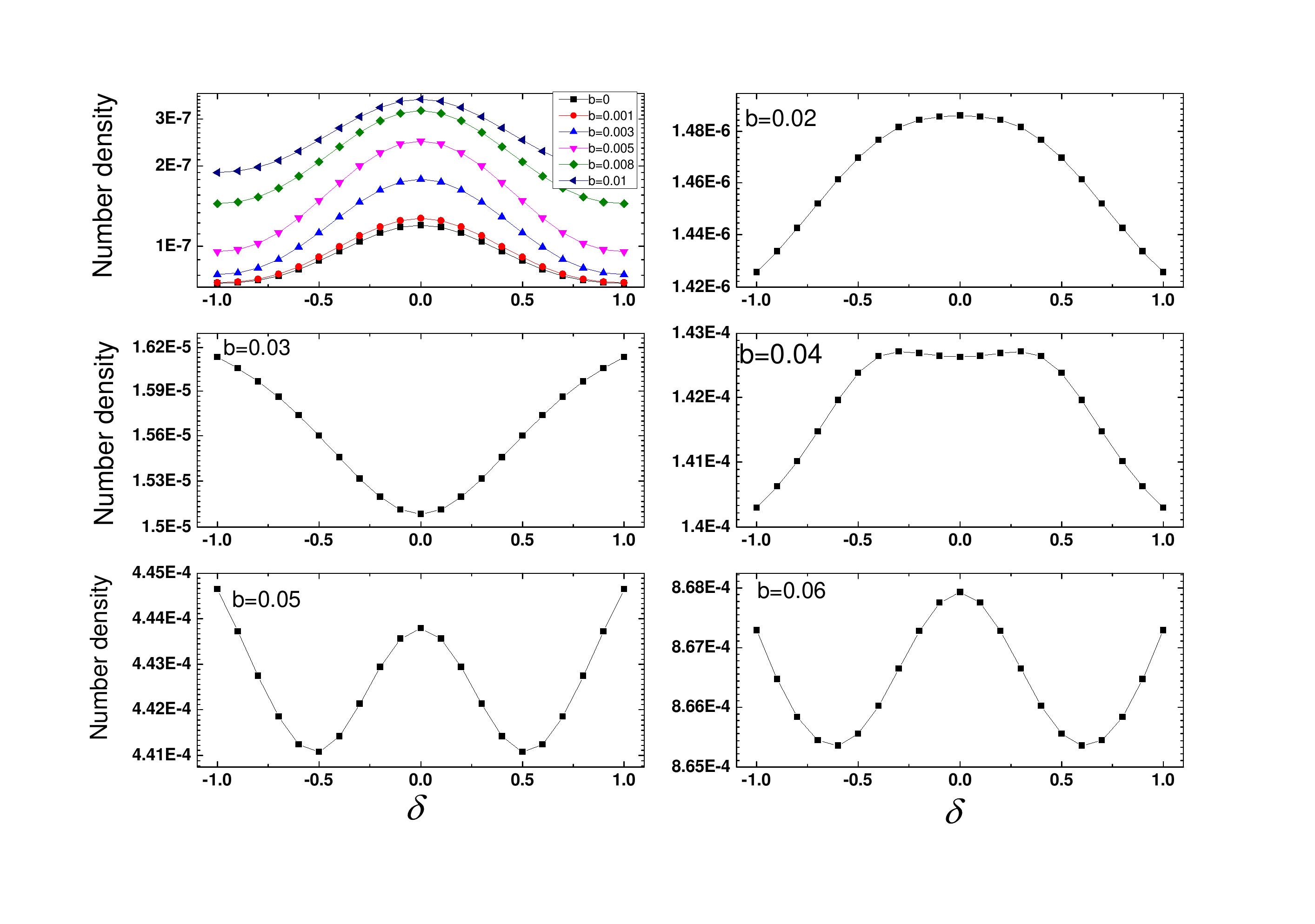}
\vspace{-27mm}
\caption{The number density (in units of $\lambda_{c}^{-3}=m^3$) of created particles as
a function of the field polarization $\delta$ for different chirp parameters $b$.
The other field parameters are the same as in Fig. \ref{fig:1}.}
\label{fig:2}
\end{figure}

The number densities as a function of the polarization parameter $\delta$ are shown in Fig. \ref{fig:2}.
They clearly display the expected symmetry when mirroring $\delta \to -\delta$. This then
implies that the case of linear polarization, $\delta=0$, provides an extremum in the number density.
As one immediately sees from  Fig. \ref{fig:2} the maximum which is present at small chirps becomes
a minimum at larger values of $b$ which then turns into a maximum again for even larger $b$-values.
Additional extrema appear for very large values, $b\ge 0.04$~$m^2$. However, the more important
effects are the following two: First, with increasing chirp the relative variation in the number density
becomes much smaller. For vanishing chirp the ratio of the number density for linear polarization to
the one for circular polarization is more than a factor of two. At $b=0.06$~$m^2$ the largest number
density deviates from the smallest one (assumed at $\delta \approx \pm 0.6$) by less than three
per mille. Second, with increasing chirp the peak number density increases significantly. This effect
is most pronounced when increasing from $b=0.02$~$m^2$ to $b=0.03$~$m^2$ for which the
number density increases by more than a factor of ten for all polarizations.

\begin{figure}[h]
\begin{center}
\includegraphics[width=0.5\textwidth]{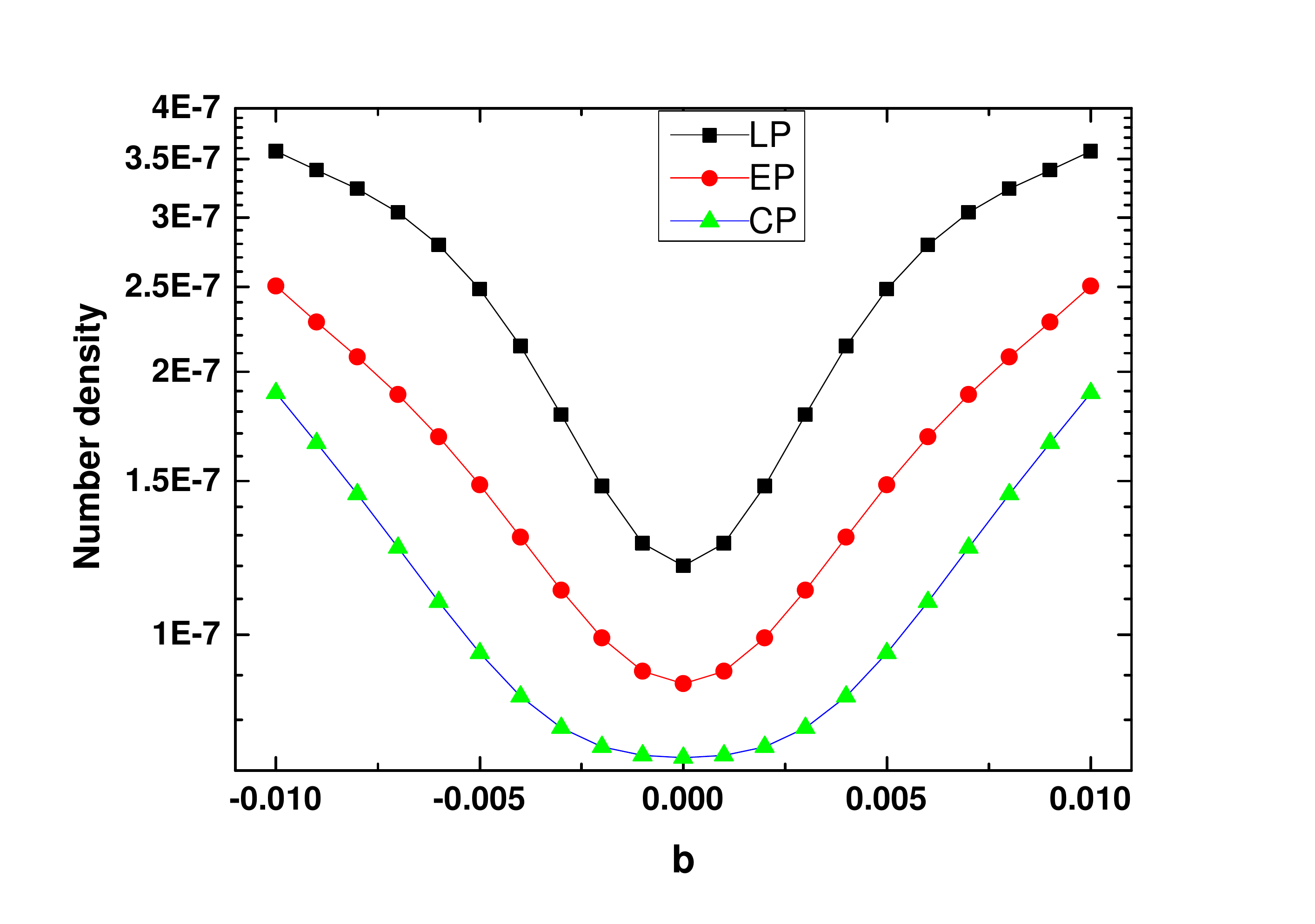}\\
~~~~~~~\includegraphics[width=0.5\textwidth]{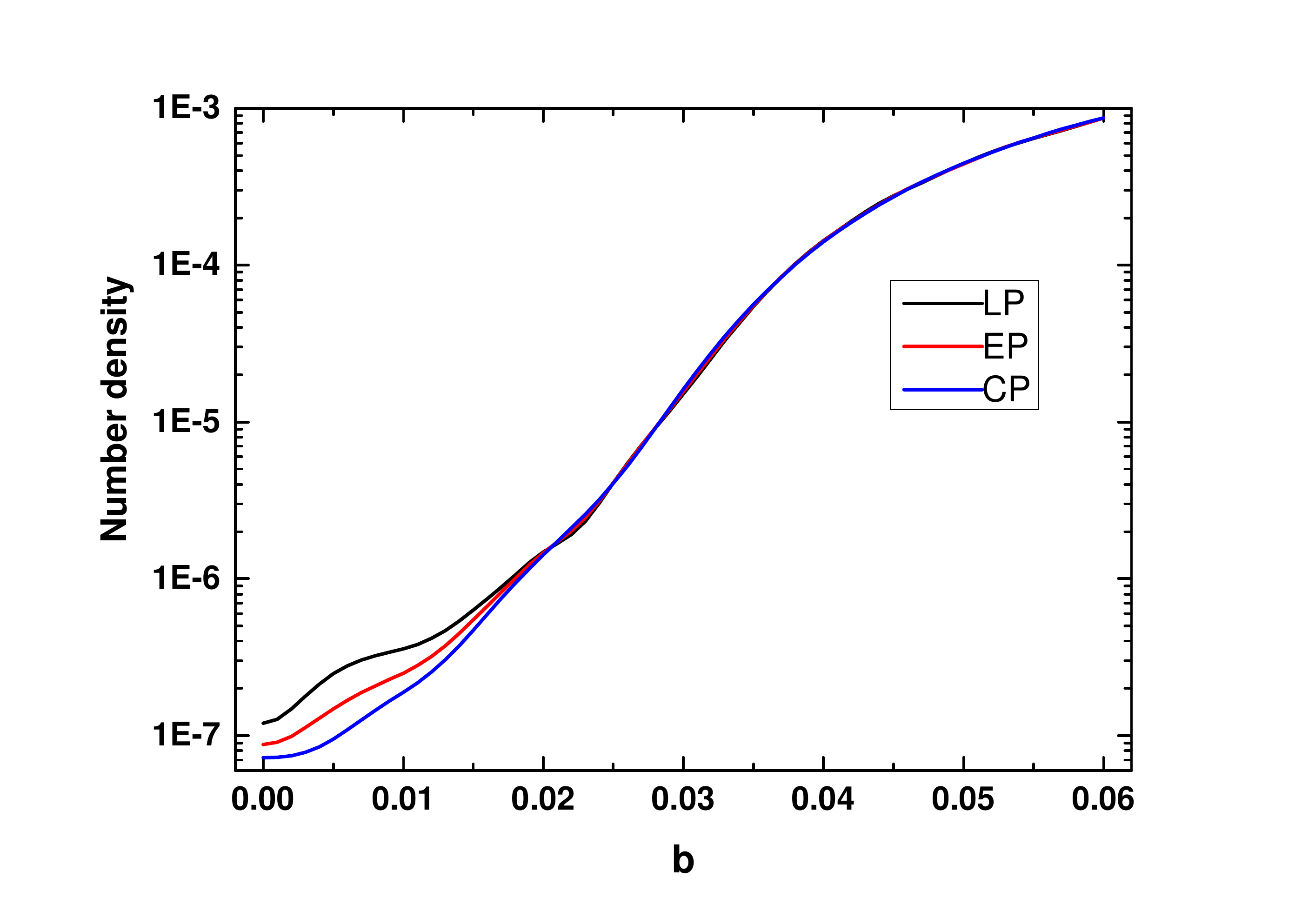}
\end{center}
\vspace{-15mm}
\caption{The number density (in units of $\lambda_{c}^{-3}=m^3$) of created particles as
a function of the chirp parameters $b$ for different polarizations $\delta=0$(LP), $\delta=0.5$(EP),
and $\delta=1$(CP), respectively. The other field parameters are the same as in Fig. \ref{fig:1}.}
\label{fig:3}
\end{figure}

When plotting the number densities as a function of the chirp parameter $b$ for the three
different polarizations $\delta=0$, $0.5$ and 1 one sees for relatively small chirp values a
symmetry under the reflection $b \to -b$, see the upper panel of Fig.~\ref{fig:3}. Also very
clearly visible is the suppression in the number densities when one goes from linear to circular
polarization. From the lower panel of Fig.~\ref{fig:3} one can infer the more or less exponential
increase in number density for increasing chirp as well as the fact that the number densities
for different polarization become degenerate for different polarizations.
As the effective frequency $\omega_{\mathrm {eff}} = \omega + b t$ increases towards the
end of the pulse, see Fig.~\ref{fig:1}, a related increase in the production rate is expected.
Nevertheless, the size of the effect is surprisingly large.
Some corresponding numbers are provided in Table \uppercase\expandafter{\romannumeral1}.

\begin{table}[ht]
\caption{Numerical results for the number densities (in units of $\lambda_{c}^{-3}=m^3$)
for some selected chirp (in units of $m^2$) and polarization parameters.}
\centering
\begin{ruledtabular}
\begin{tabular}{ccccc}
$n$ & b= 0& b=0.02 & b = 0.04 & b=0.06 \\
\hline
$\delta=0$    &$1.200\times10^{-7}$ & $1.486 \times 10^{-6}$ & $1.426 \times 10^{-4}$  & $0.8679\times10^{-3}$\\
\hline
$\delta=0.5$ &$0.880\times10^{-7}$ &  $1.470 \times 10^{-6}$ & $1.424 \times 10^{-4}$  & $0.8656\times10^{-3}$\\
\hline
$\delta=0.9$ &$0.728\times10^{-7}$ &  $1.434 \times 10^{-6}$ & $1.406 \times 10^{-4}$   & $0.8665\times10^{-3}$\\
\hline
$\delta=1$    &$0.724\times10^{-7}$ &  $1.426 \times 10^{-6}$ & $1.403 \times 10^{-4}$  & $0.8673\times10^{-3}$\\
\end{tabular}
\end{ruledtabular}
\vskip12pt
\end{table}

Here, a technical remark is in order. For some large values of the chirp parameter $b$,
{\it e.g.}, around $b=0.05$~$m^2$, some irregular changes of the numerical results for
the number densities for each polarization are observed. As these have to be very likely
attributed to instabilities of the numerical procedure we are refraining from displaying these
exceptional points here. Nevertheless, this issue will be clarified in future investigations.


\section{Momentum spectra}\label{result2}

In this section, we will report on results for the momenta spectra (MS)  of the
produced particles for several chirp parameters each for the cases of
(i) linear polarization ($\delta=0$), (ii)  elliptical polarization ($\delta=0.5$),
(iii)  near-circular elliptical polarization ($\delta=0.9$), and
(iv) circular polarization ($\delta=1$).

\subsection{Momentum spectra for linear polarization $\delta=0$}
\label{SecLinPol}

\begin{figure}[ht]
\begin{center}
\includegraphics[width=\textwidth]{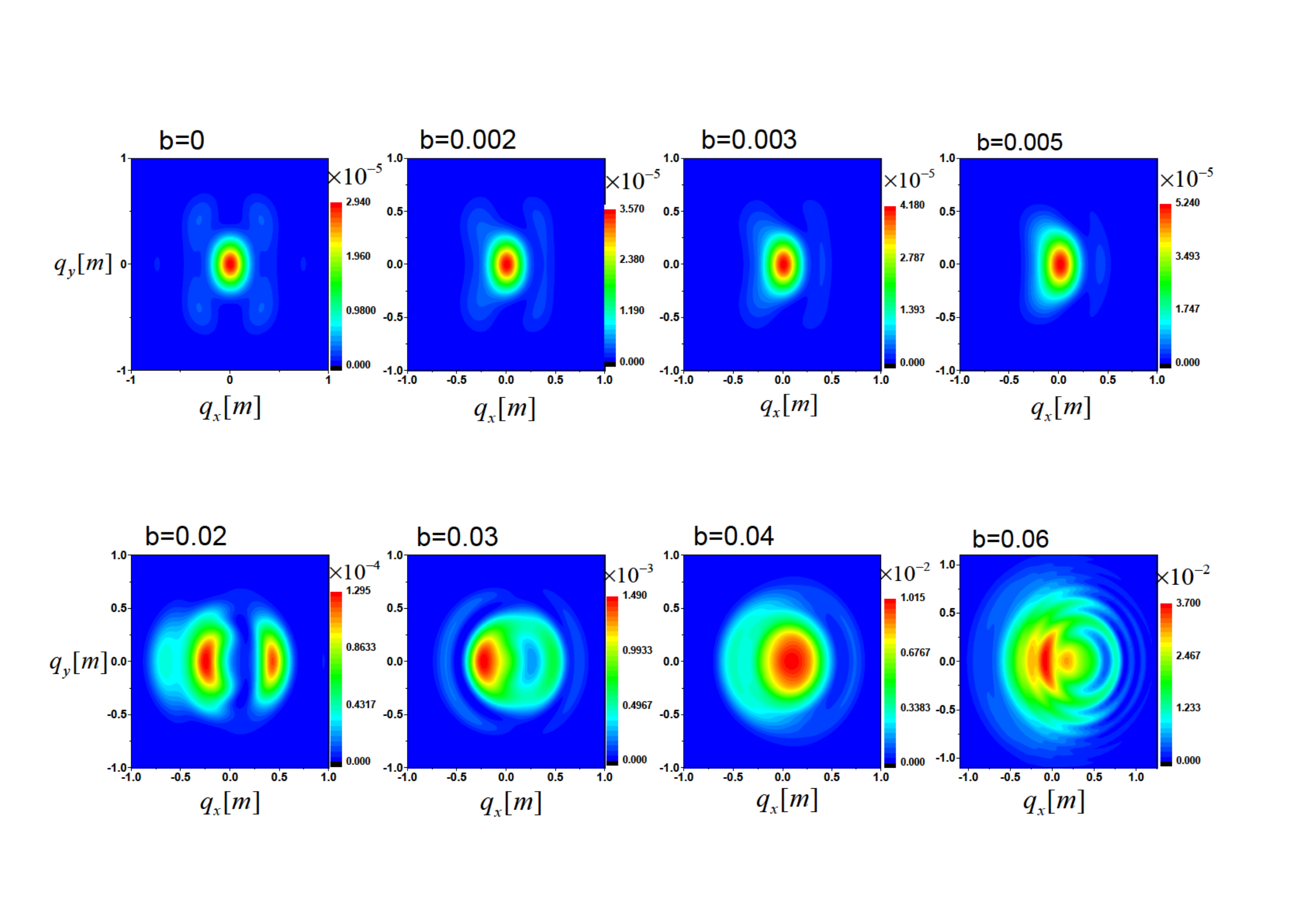}
\end{center}
\vspace{-15mm}
\caption{Momentum spectra  of produced $e^{+}e^{-}$ pairs for linear polarization ($\delta=0$)
at $q_z=0$ in the $( q _x,q_y)$-plane. The other field parameters are given in Eq.~\eqref{FieldParameters}.
Upper row: from left to right the values of the small chirp parameters are
$b$=0, 0.002$m^2$, 0.003$m^2$ and 0.005$m^2$, respectively.
Lower row: from left to right the values of the large chirp parameters are $b$=0.02$m^2$, 0.03$m^2$,
 0.04$m^2$ and 0.06$m^2$, respectively.}
\label{fig:4}
\end{figure}

First of all, we note that in the case of linear polarization the electric field is oriented only
along the $x$-axis, and the momentum spectra possess correspondingly a rotational
symmetry around the $q_x$-axis. For the linear polarized $(\delta=0)$ pulse
the momentum spectra in $q_x$ and $q_y$
for $q_z=0$ are plotted in Fig.~\ref{fig:4}.. For vanishing chirp,
$b=0$, the
results agree with the ones of a previous investigation \cite{Li:2017qwd}.
For non-vanishing chirp parameters the main
result is, besides the expected lower symmetry of the spectra,
the appearance of strong interference effects  leading to several
maxima and minima of the pair production rate as a function of momenta.

As can be seen in the upper panel of Fig.~\ref{fig:4} small chirp parameters lead to
small variations in the spectrum: There is a slight enhancement in the height of the peak,
a small shift towards positive $q_x$, and a broadening as well as the loss of one of
the reflection symmetries of the peak.

For larger frequency chirps, $b\geq 0.02m^2$, the momentum spectra display some remarkable
structures as can be seen from the lower row in Fig.~\ref{fig:4}. For $b$=0.02$m^2$ the spectrum
possesses two peaks at negative and positive $q_{x}$, respectively. For $b$=0.03$m^2$ the
main peak is located at negative $q_{x}$.  For $b$=0.04$m^2$ the peak goes back to positive
momenta, and at $b$=0.06$m^2$ again to negative momenta. In the latter case strong interference
effects are visible, note especially the ring-like structure.

\begin{figure}[ht]
\begin{center}
\includegraphics[width=\textwidth]{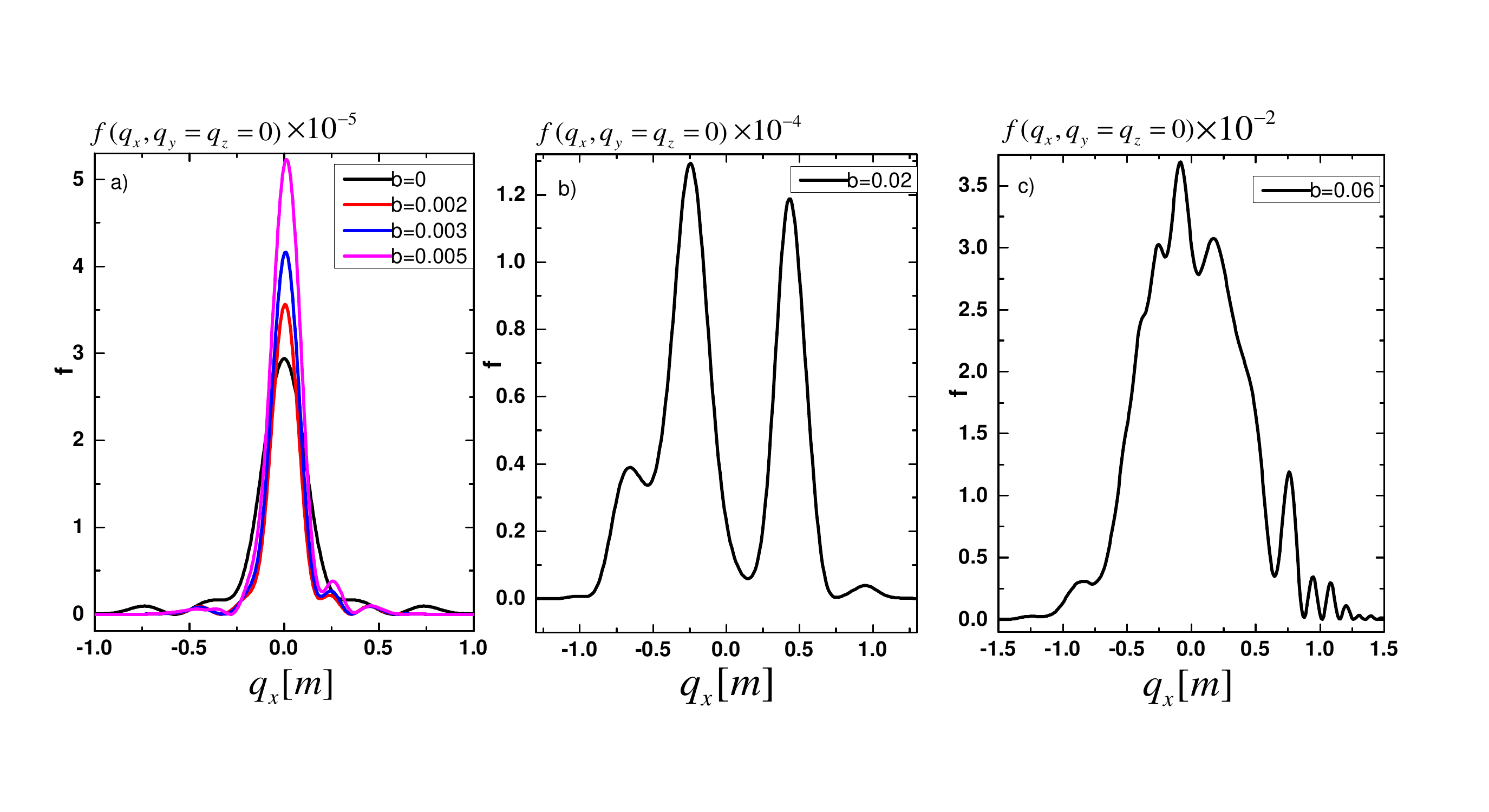}
\end{center}
\vspace{-15mm}
\caption{Momentum spectra  of produced pairs for linear polarization ($\delta=0$)
at $q_y=q_z=0$, {\it cf}.\ Fig.~\ref{fig:4}. }
\label{fig:5}
\end{figure}

In all the cases shown in Fig.~\ref{fig:4} the spectra are symmetric w.r.t. to reflection of $q_y$.
(NB: For non-vanishing $q_z$ the spectra would be symmetric w.r.t. to a rotation around the
$q_x$-axis.)
Therefore, the effects of chirps can be understood in more detail by plotting the number density
for $q_y=q_z=0$ as a function of $q_x$. In the left panel of Fig.~\ref{fig:5}(a)  one sees that
for vanishing chirp the peak is located at $q_{x}=0$. As $b$ increases, the peak is very slightly
shifted to positive $q_{x}$. For $b$=0.02$m^2$  one sees now not only the two prominent
maxima but also additional but less pronounced ones.  At $b$=0.06$m^2$ the momentum
spectrum displays quite complicated interference patterns but note also the change in
the height of the peaks, enhanced from $2.94\times10^{-5}$ ($b=0$) to $3.7\times10^{-2}$
( $b=0.06m^2$).

As we will argue in  Sec.~\ref{result3} the effects of the frequency chirp on the spectra can be
explained by a semi-classical analysis based on the WKB approximation. As the related
effective potential changes with the frequency chirp the numerically observed drastic
effects are plausible.

\subsection{Elliptical polarization $\delta=0.5$}

\begin{figure}[h]
\begin{center}
\includegraphics[width=\textwidth]{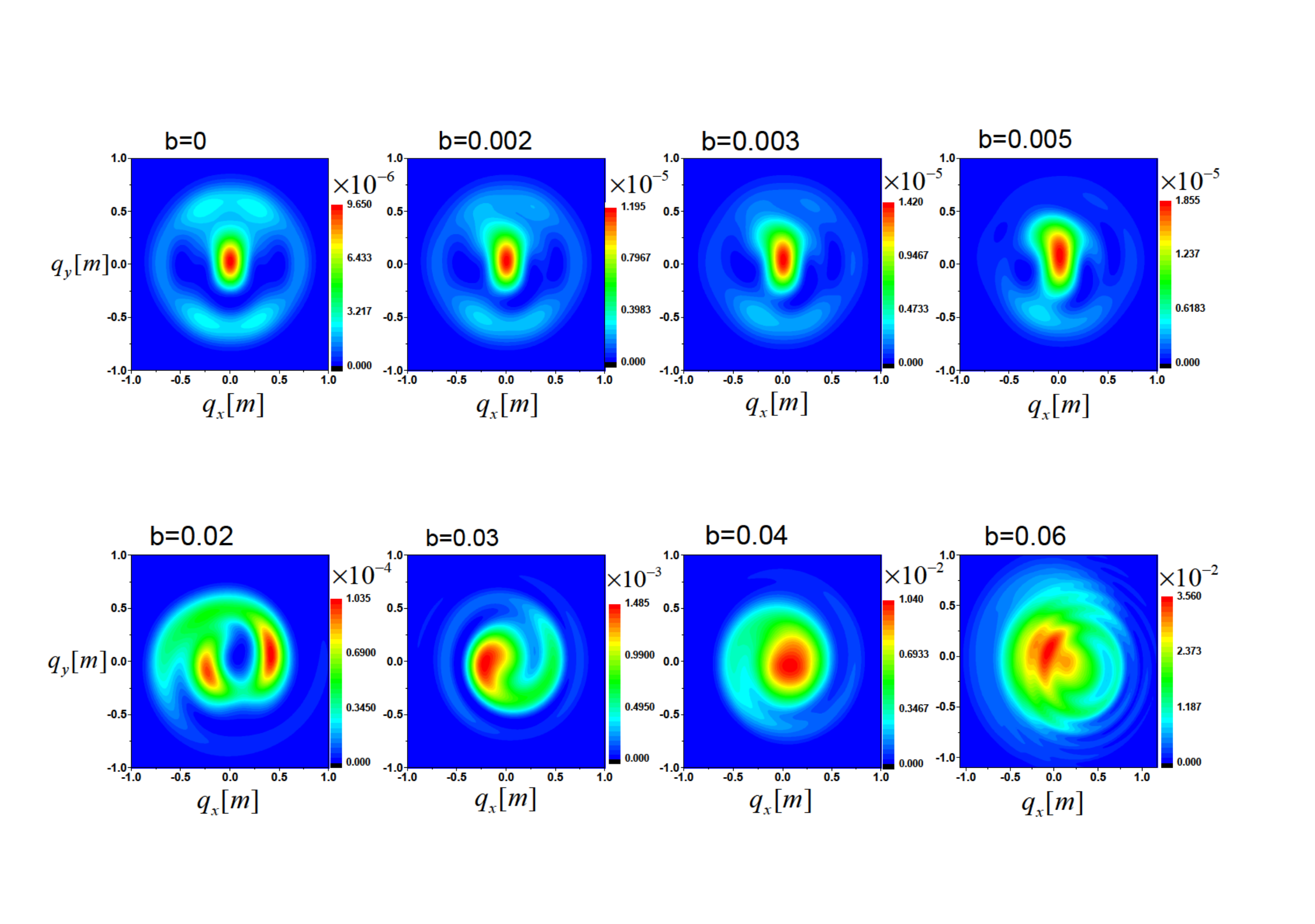}
\end{center}
\vspace{-15mm}
\caption{Momentum spectra  of produced $e^{+}e^{-}$ pairs for elliptic polarization ($\delta=0.5$)
at $q_z=0$ in the $( q _x,q_y)$-plane. The other field parameters are given in Eq.~\eqref{FieldParameters}.
Upper row: from left to right the values of the chirp parameters are
$b$=0, 0.002$m^2$, 0.003$m^2$ and 0.005$m^2$, respectively.
Lower row: from left to right the values of the chirp parameters are $b$=0.02$m^2$, 0.03$m^2$,
 0.04$m^2$ and 0.06$m^2$, respectively.}
\label{fig:6}
\end{figure}

As a next case we consider the momentum spectra for an elliptically polarized electric field,
$\delta=0.5$., see Fig.~\ref{fig:6}. Whereas for $b=0$ there is still a reflection symmetry around
the $q_{x}$-axis also this symmetry gets lost when $b\not=0$. For small chirp parameters
the distortion of the spectrum is again quite mild, see the upper panel of Fig.~\ref{fig:6}.
Quite some complicated reordering of the spectra take place for large values
of the chirp parameter, {\it cf.}, the lower panel of Fig.~\ref{fig:6}.
Especially the splitting into several extrema is very similar to what happens in
the linear polarized case.

\subsection{Near-circular elliptic polarization $\delta=0.9$}

\begin{figure}[ht]
\begin{center}
\includegraphics[width=\textwidth]{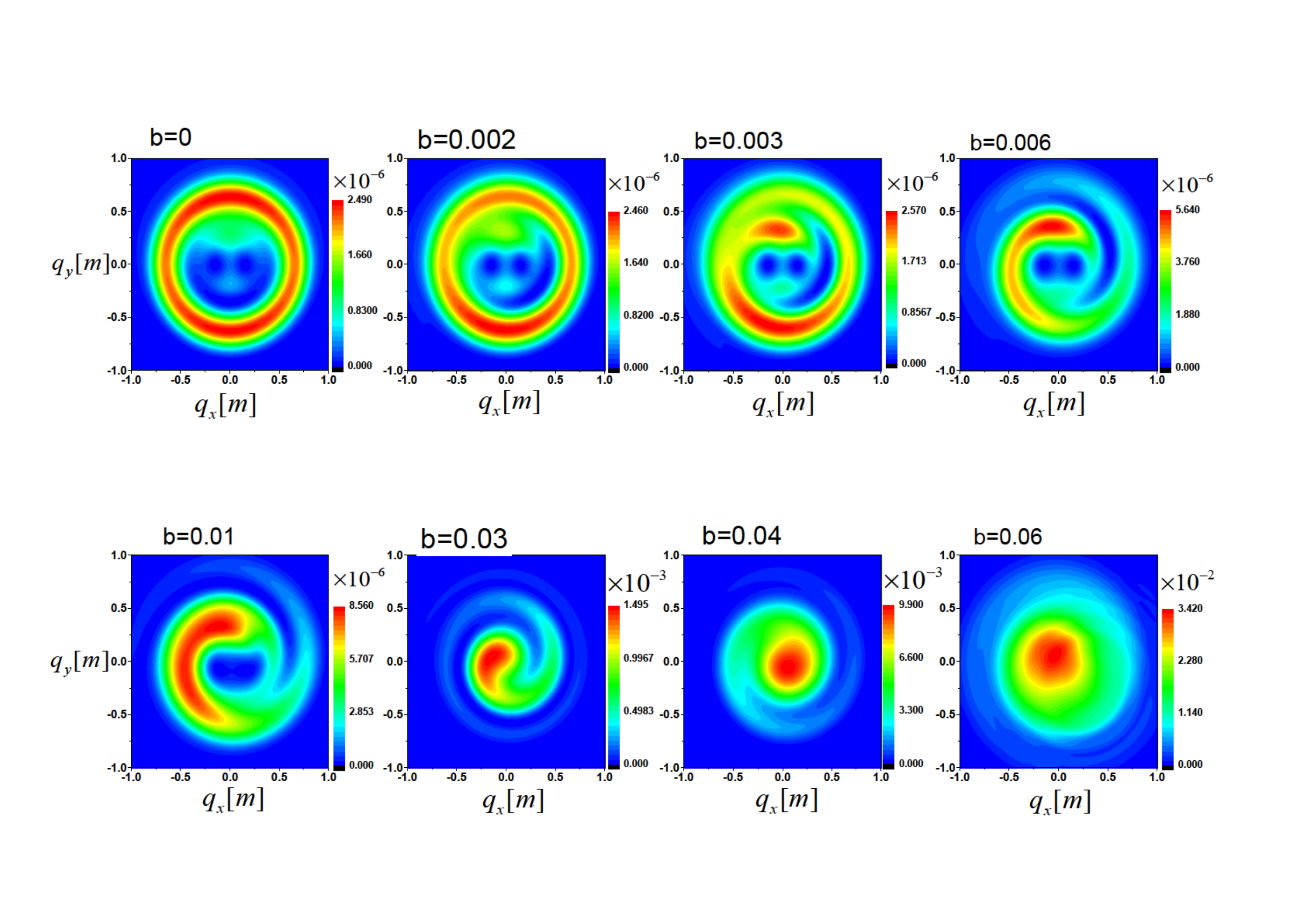}
\end{center}
\vspace{-15mm}
\caption{Momentum spectra  of produced $e^{+}e^{-}$ pairs for
near-circular elliptic polarization ($\delta=0.9$)
at $q_z=0$ in the $( q _x,q_y)$-plane. The other field parameters are given in Eq.~\eqref{FieldParameters}.
Upper row: from left to right the values of the chirp parameters are
$b$=0, 0.002$m^2$, 0.003$m^2$ and 0.006$m^2$, respectively.
Lower row: from left to right the values of the chirp parameters are $b$=0.01$m^2$, 0.02$m^2$,
 0.03$m^2$ and 0.06$m^2$, respectively.}
\label{fig:7}
\end{figure}

For the near-circular elliptically polarized case we plot the spectra
in Fig.~\ref{fig:7}.
For $b=0$ the main peak region is ring-shaped.
This can be understood from the fact that the electric field changes
its direction  during the pair creation process.
Thus, the particles may be accelerated into different directions depending on
the field direction at the time of production. These findings are very
similar to the results of the strong-field ionization of helium using an
elliptically polarized laser pulses \cite{Pfeiffer}
and the effects of electric field polarizations on pair production from
 vacuum \cite{Li:2015cea}.

Otherwise, one sees also drastic effects of the chirp for relatively small chirp
parameters, and especially the ring form is distorted to  a spiral one.
For very large chirp parameters, on the other hand, the spectrum shows
in this case less structure.

In contrast to the previously discussed linear polarization in the case of the near-circular elliptic polarization
the spectra loose their symmetry in both of $q_{x}$ and $q_{y}$. The patterns observed in
 Fig.~\ref{fig:7} will become clearer when discussing the case of perfect circular polarization.

\subsection{Circular polarization $\delta=1$}

\begin{figure}[ht]
\begin{center}
\includegraphics[width=\textwidth]{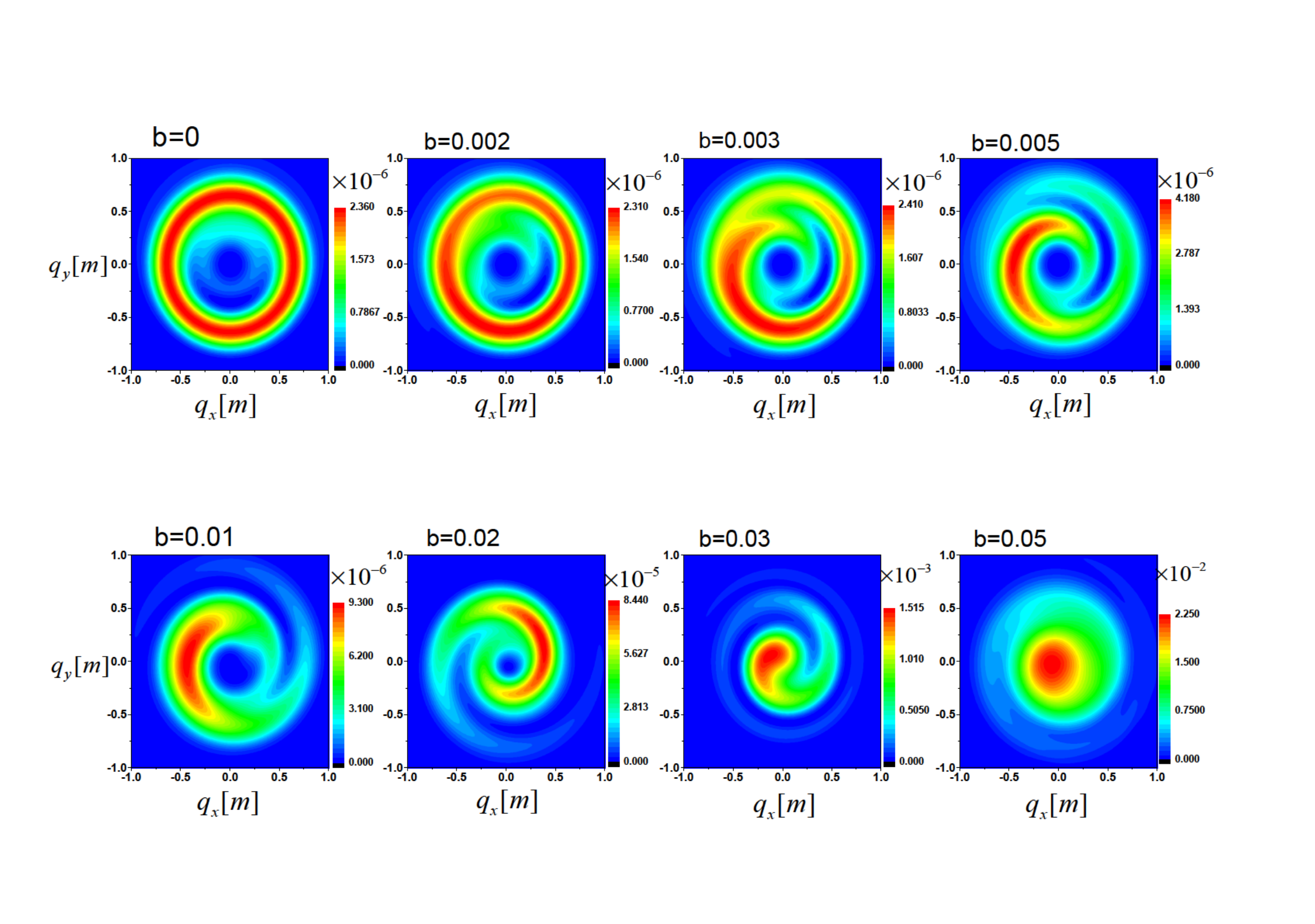}
\end{center}
\vspace{-15mm}
\caption{Momentum spectra  of produced $e^{+}e^{-}$ pairs for
circular polarization ($\delta=1$)
at $q_z=0$ in the $( q _x,q_y)$-plane. The other field parameters are given in Eq.~\eqref{FieldParameters}.
Upper row: from left to right the values of the chirp parameters are
$b$=0, 0.002$m^2$, 0.003$m^2$ and 0.005$m^2$, respectively.
Lower row: from left to right the values of the chirp parameters are $b$=0.01$m^2$, 0.02$m^2$,
 0.03$m^2$ and 0.05$m^2$, respectively.}
\label{fig:8}
\end{figure}

For the circularly polarized ($\delta=1$) the spectra are shown in Fig. \ref{fig:8}.
The $b=0$ spectrum, shown at the top and most left,  displays a ring-shaped
maximum centered around the origin. It
exhibits the weak interference pattern or/and oscillation between the hole and outer ring along
negative values of $q_{y}$ \cite{Blinne:2013via,Li:2015cea,Li:2017qwd}.
This might be interpreted within a semi-classical analysis by means of an effective
 scattering potential \cite{Dumlu:2010vv}, see the next section.
The outer ring structure results from multi-photon pair creation \cite{Blinne:2013via},
and in the strong field limit its radius can be determined by the energy conservation
to be at $| \mathbf{q} | =1/2\sqrt{(n\omega)^{2}-4m^{2}}$ where $n$ is the number
of photons participating in the pair creation, {\it cf.} ref.~\cite{Kohlfurst:2013ura}.

Again the spectra are very sensitive to chirps even for relatively small chirp parameters.
This includes the distortion of the ring structure, the appearance of spirals, and,
last but not least, a significant increase of the one-particle distribution function.
The characteristic shape of the spectra could be helpful in an experimental identification
of pair production. {\it E.g.}, at $b=0.03$ $m^2$ one clearly identifies an Archimedian spiral
which is going to start from almost the central region slightly shifted to  negative $q_{x}$ values.
For very large chirp parameters, $b\geq0.05$ $m^2$, the spiral structure is fading away,
and the spectra become less structured.

\bigskip

\begin{table}[ht]
\caption{The peak values for the one-particle distribution function at late times $f(\mathbf{q},\infty)$.
Note that these peaks occur at different values of the momentum $\mathbf{q}$, see the discussion
above.}
\centering
\begin{ruledtabular}
\begin{tabular}{lcc}
$f(\mathbf{q},\infty)$ at peak &$b=0$ & $b=0.06$ $m^2$ \\
\hline
$\delta=0$    &$29.4\times10^{-6}$   &$ 3.70\times10^{-2}$\\
\hline
$\delta=0.5$ &$9.65\times10^{-6}$   &$ 3.56\times10^{-2}$\\
\hline
$\delta=0.9$ &$2.49\times10^{-6}$   &$ 3.42\times10^{-2}$\\
\hline
$\delta=1$    &$2.36\times10^{-6}$   &$ 3.41\times10^{-2}$\\
\end{tabular}
\end{ruledtabular}
\vskip12pt
\end{table}

To summarize this section, we have obtained quite some detailed information how the
spectra of the produced pairs change for a given polarization when frequency chirps
from relatively modest to quite large ones are considered. The positions of the global
extrema of the one-particle distribution function, {\it i.e.}, the peaks of the spectra
display a quite rich structure. Common to all the considered cases is the strong increase
in peak values for increasing chirp parameters which is easily understood from
the effective frequency, $\omega_{\mathrm {eff}} = \omega + b t$, of the field,
respectively, by the onset of multi-photon pair production.
Some corresponding values are given in Table \uppercase\expandafter{\romannumeral2}.

\section{Semi-classical analysis}\label{result3}

In this section we will employ a semi-classical analysis to obtain a qualitative
explanation of the effects of chirps on the spectrum. Hereby we follow the
WKB method outlined in ref.~\cite{Akkermans:2011yn}. These authors
considered a Sauter pulse, {\it i.e.}, a gauge potential $A(t) \propto \tanh \omega t$,
for which the turning points of the analogue semi-classical scattering potential
can be determined analytically by solving the condition
$\Omega(\mathbf{q},t_{p})= \sqrt{m^{2}+(\mathbf{q}-e\mathbf{A}(t_p))^{2}}=0$.
In this case a single complex-conjugated pair dominate, and the spectra
of the produced particles can be qualitatively explained (see
Fig. 3 in Ref. \cite{Akkermans:2011yn}).
The WKB result for the created number of pairs of momentum $\mathbf q$ is hereby given by
\begin{equation}\label{4}
  N_{q}\approx \exp(-2K) \, , \quad
 K =  \left| \int^{t_{2}}_{t_{1}}  dt \Omega(\mathbf{q},t)dt \right| \, ,
\end{equation}
where $t_{1}$ and $t_{2}$ are dominant turning points closed to the real $t$ axis.

In case of $A(t) \propto 1 / (\omega(1+\omega^{2}t^{2}))$ one obtains two pairs of complex
turning points  which then explain interference effects \cite{Dumlu:2010ua}.
The production rate was then estimated to be a sum of two terms which takes the
form
\begin{equation}\label{5}
  N_{q}\approx e^{-2K_{1}}+e^{-2K_{2}}\pm2\cos(2\alpha)e^{-K_{1}-K_{2}},
\end{equation}
where the $+$ and $-$ signs refer to bosonic and fermionic pair production, respectively.
Hereby, the $K_{1,2}$ are obvious generalizations of the $K$ above, and $\alpha$ represents
some respective phase, for details see ref.~\cite{Dumlu:2010ua}.
The interference term  in Eq. (\ref{5}) is responsible for the oscillations in the spectra.
As the integrals $K$, respectively, $K_{1,2}$,  are almost linear in $t$ the dominant
contribution originates from the terms involving turning points which are closest to the real $t$ axis.

\begin{figure}
\centering
  \begin{minipage}{7cm}
   \includegraphics[width=7cm]{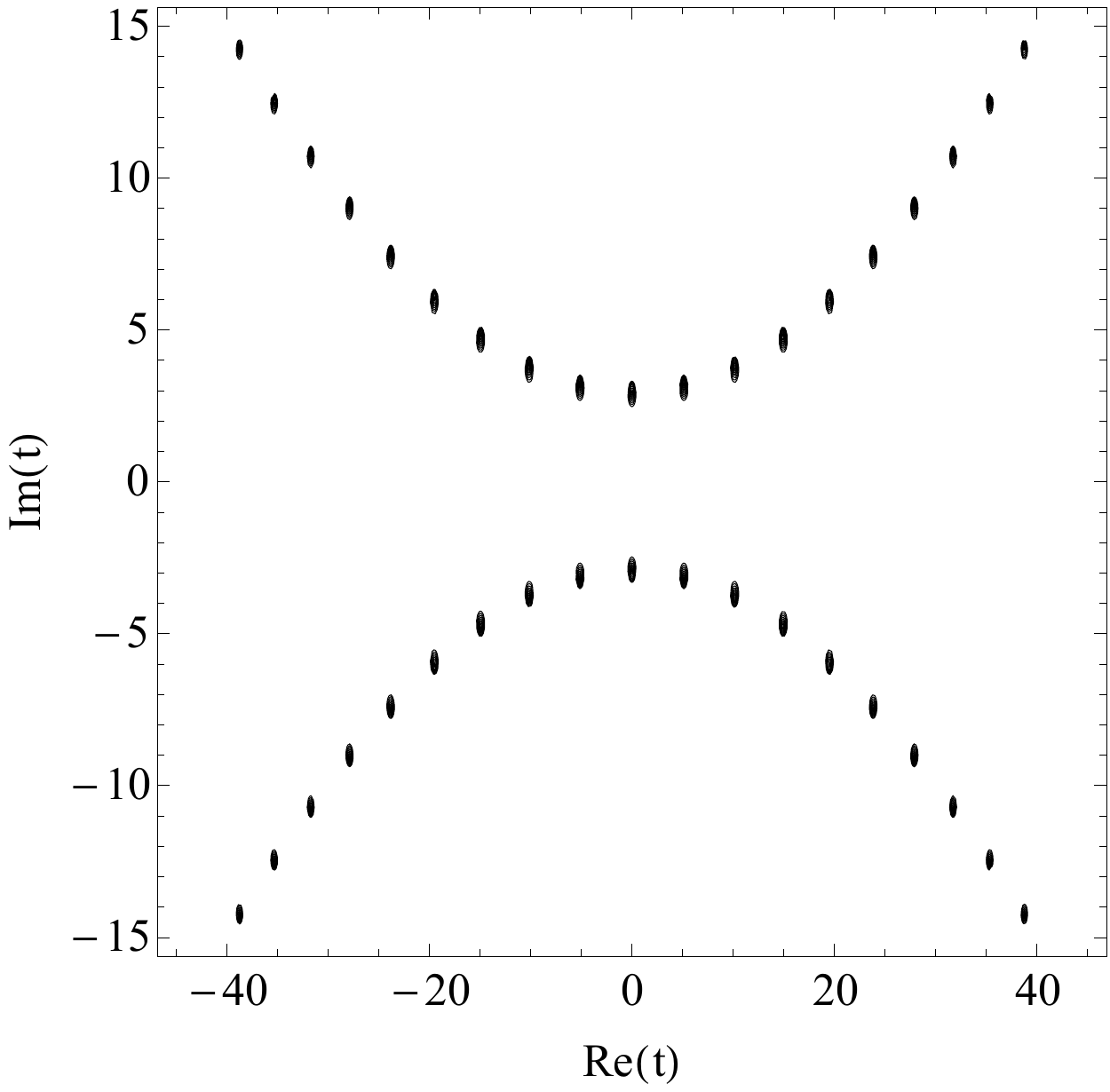}
  \end{minipage}
  \begin{minipage}{7cm}
    \includegraphics[width=7cm]{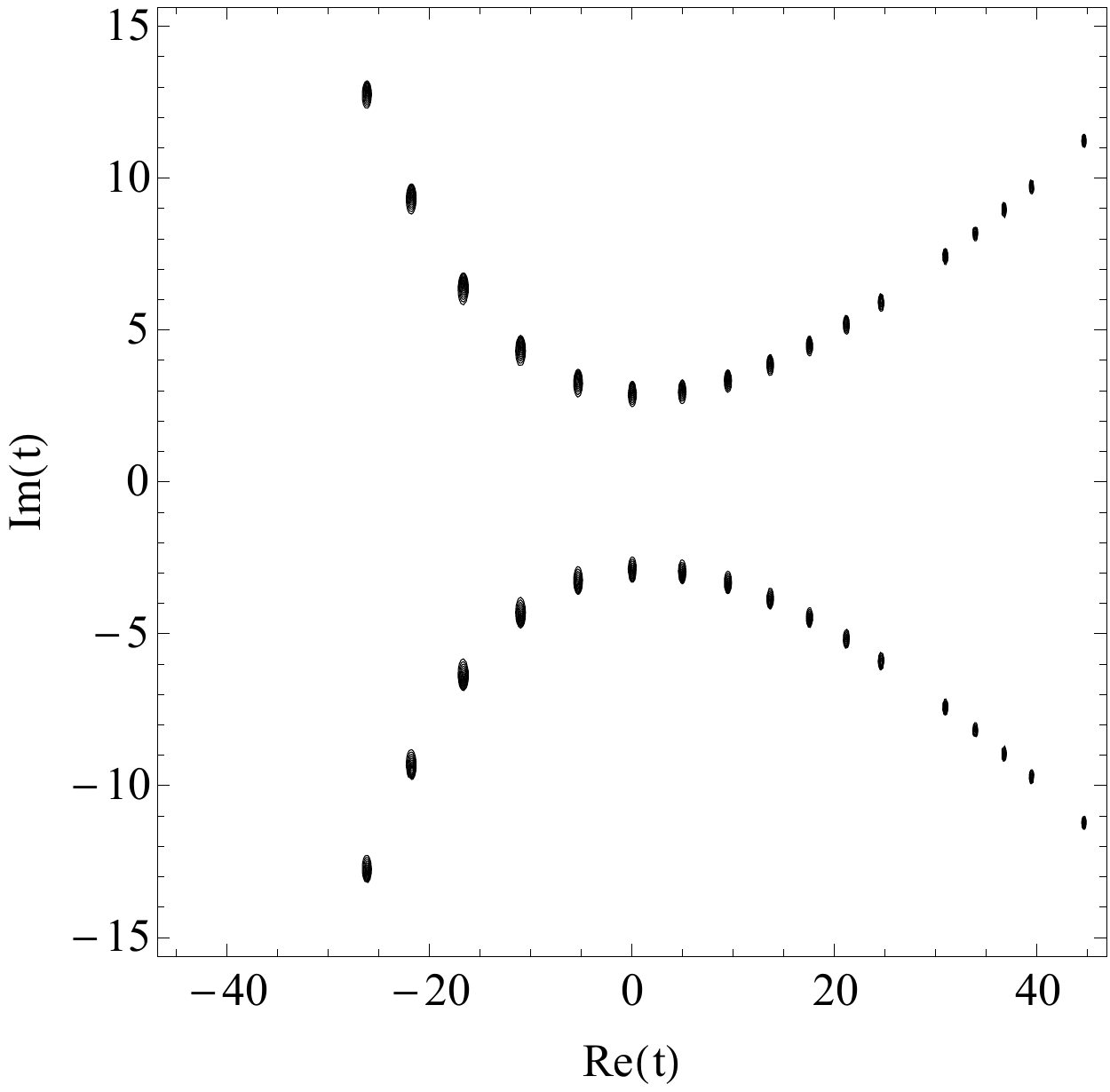}
  \end{minipage}

  \begin{minipage}{7cm}
   \includegraphics[width=7cm]{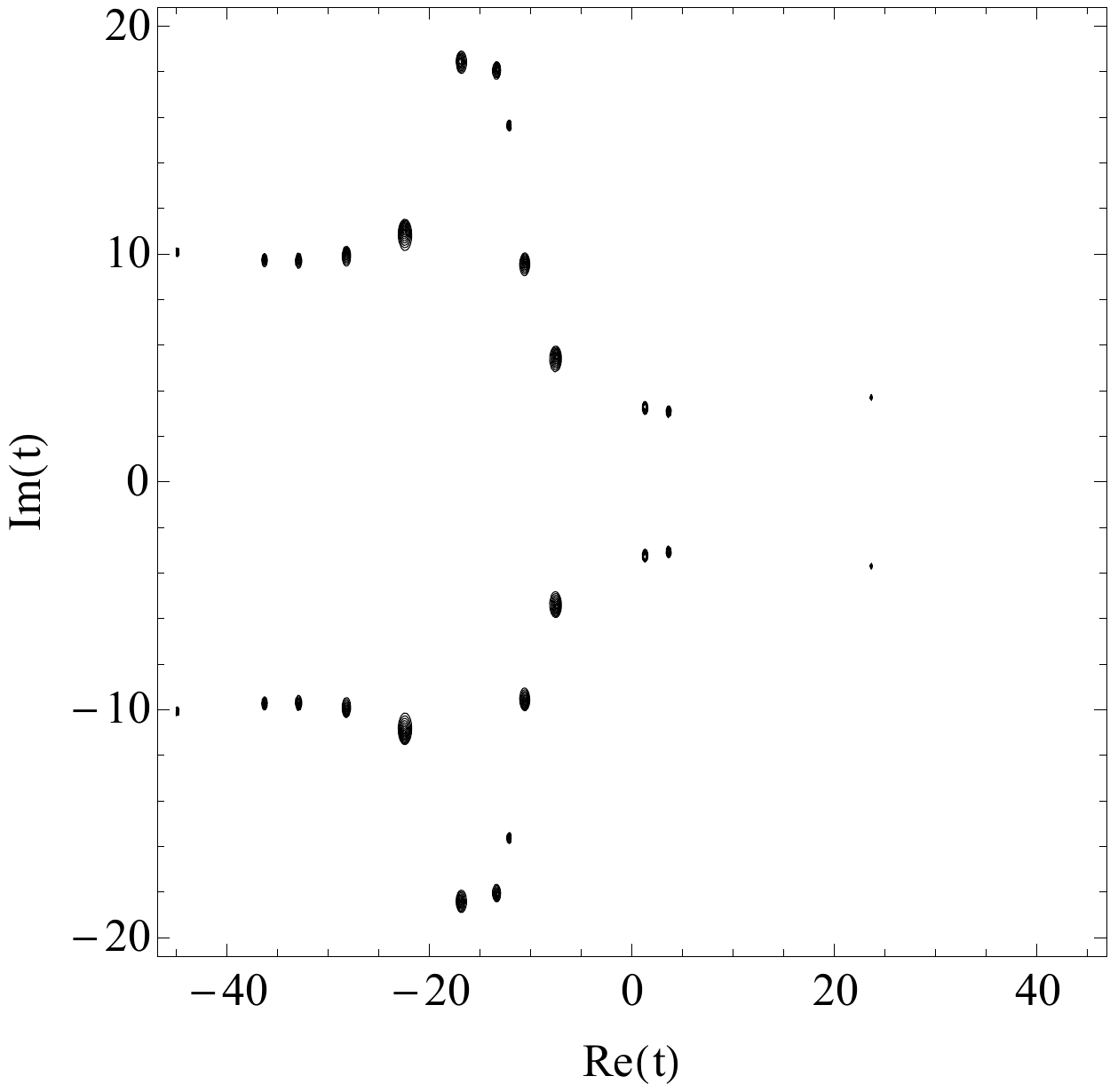}
  \end{minipage}
  \begin{minipage}{7cm}
   \includegraphics[width=7cm]{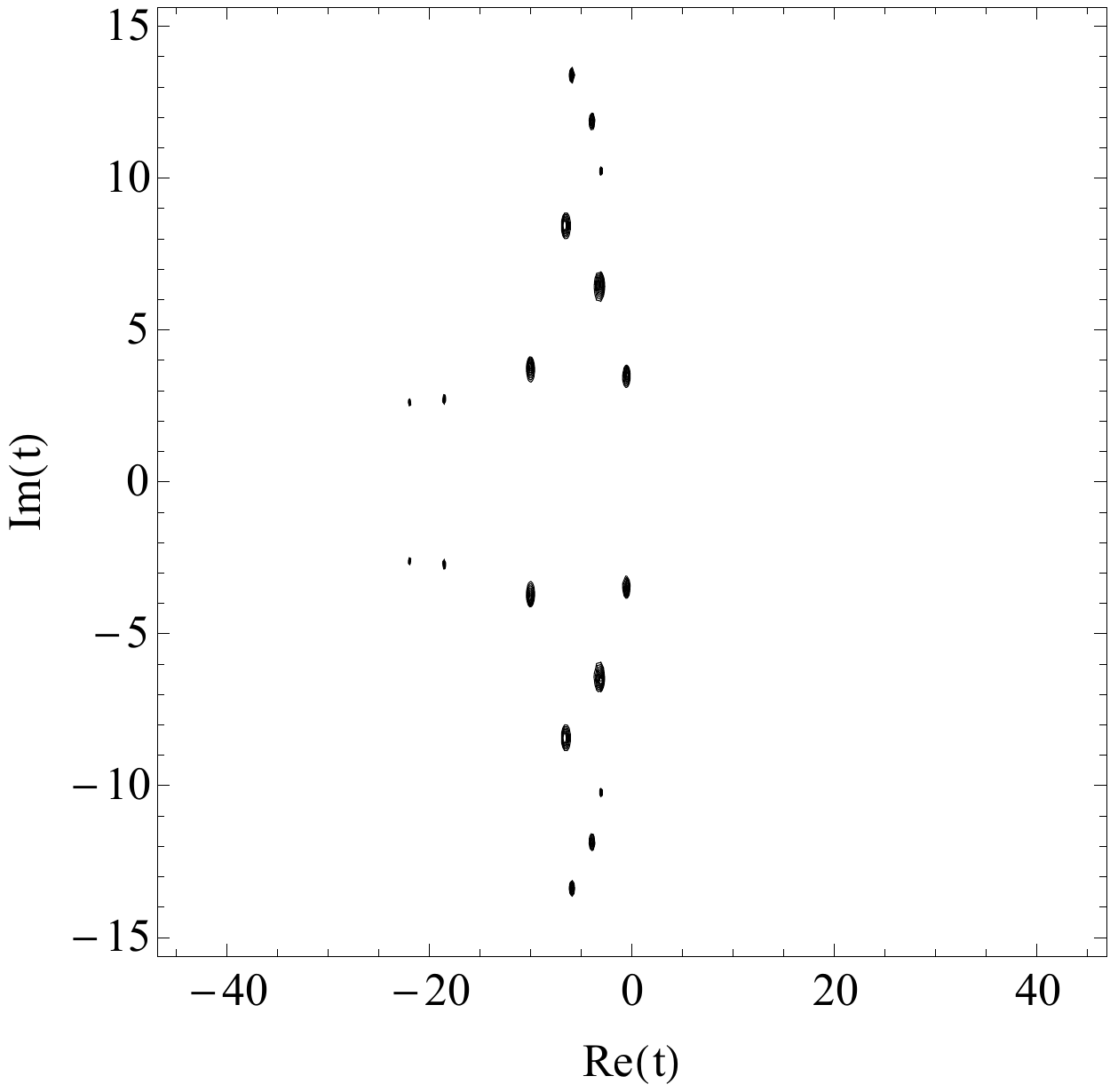}
  \end{minipage}
  \caption{Contour plots of $|\Omega (\mathbf{q},t)|^{2}$ in the complex $t$ plane,
  showing the location of turning points where $\Omega (\mathbf{q},t)=0$.
  These plots are for the linear polarization $\delta=0$. The other field parameters are the
  same as in Fig.~\ref{fig:1}.
  From top left to bottom right the values of chirp parameters are $b=0, 0.005, 0.02, 0.06$ $m^2$,
  respectively,
  and momentum values  are $q=0, 0, -0.1, -0.18$ $m$, respectively.}
  \label{fig:9}
\end{figure}

\begin{figure}
\centering
  \begin{minipage}{7cm}
   \includegraphics[width=7cm]{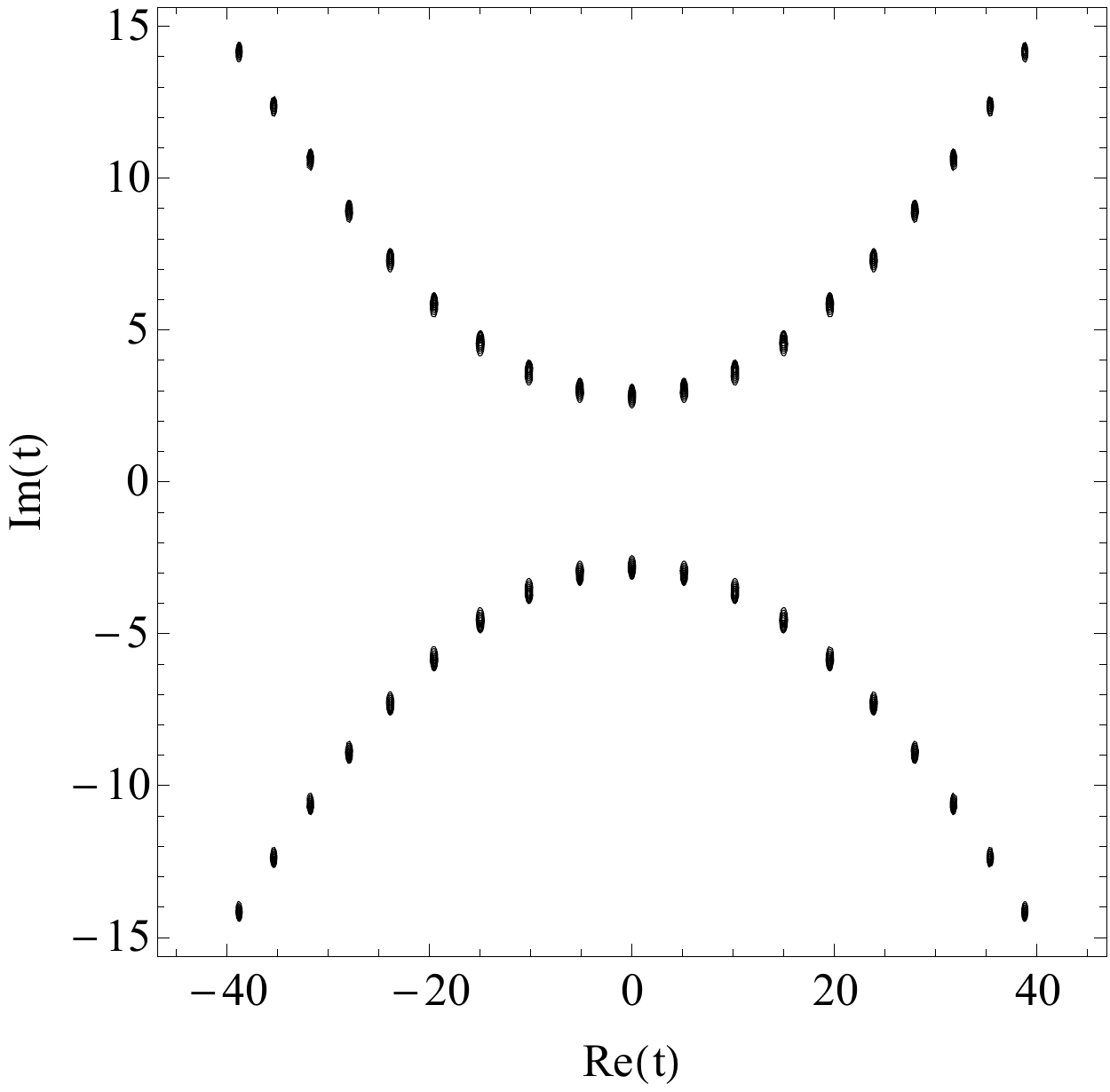}
  \end{minipage}
  \begin{minipage}{7cm}
    \includegraphics[width=7cm]{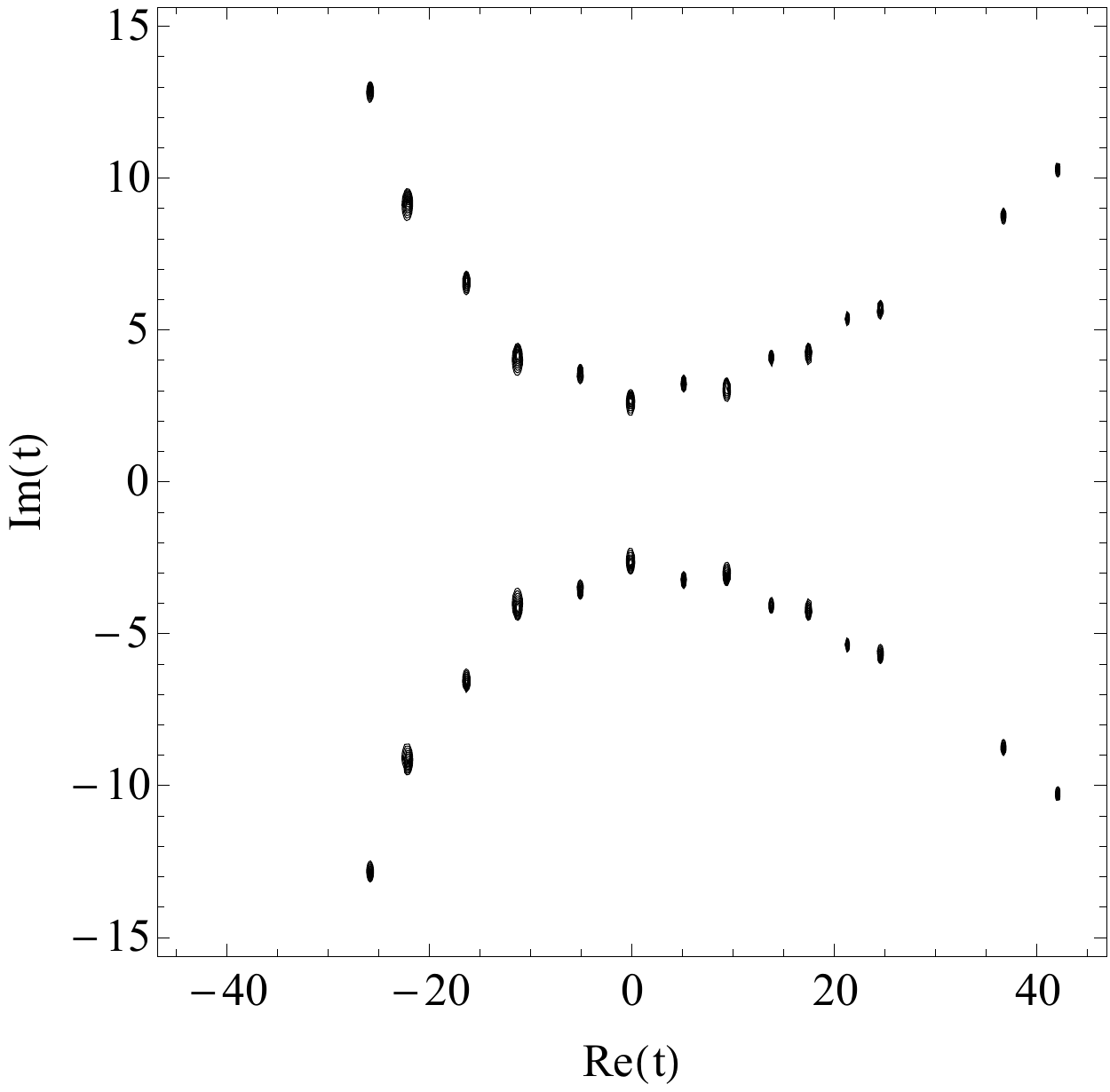}
  \end{minipage}

  \begin{minipage}{7cm}
   \includegraphics[width=7cm]{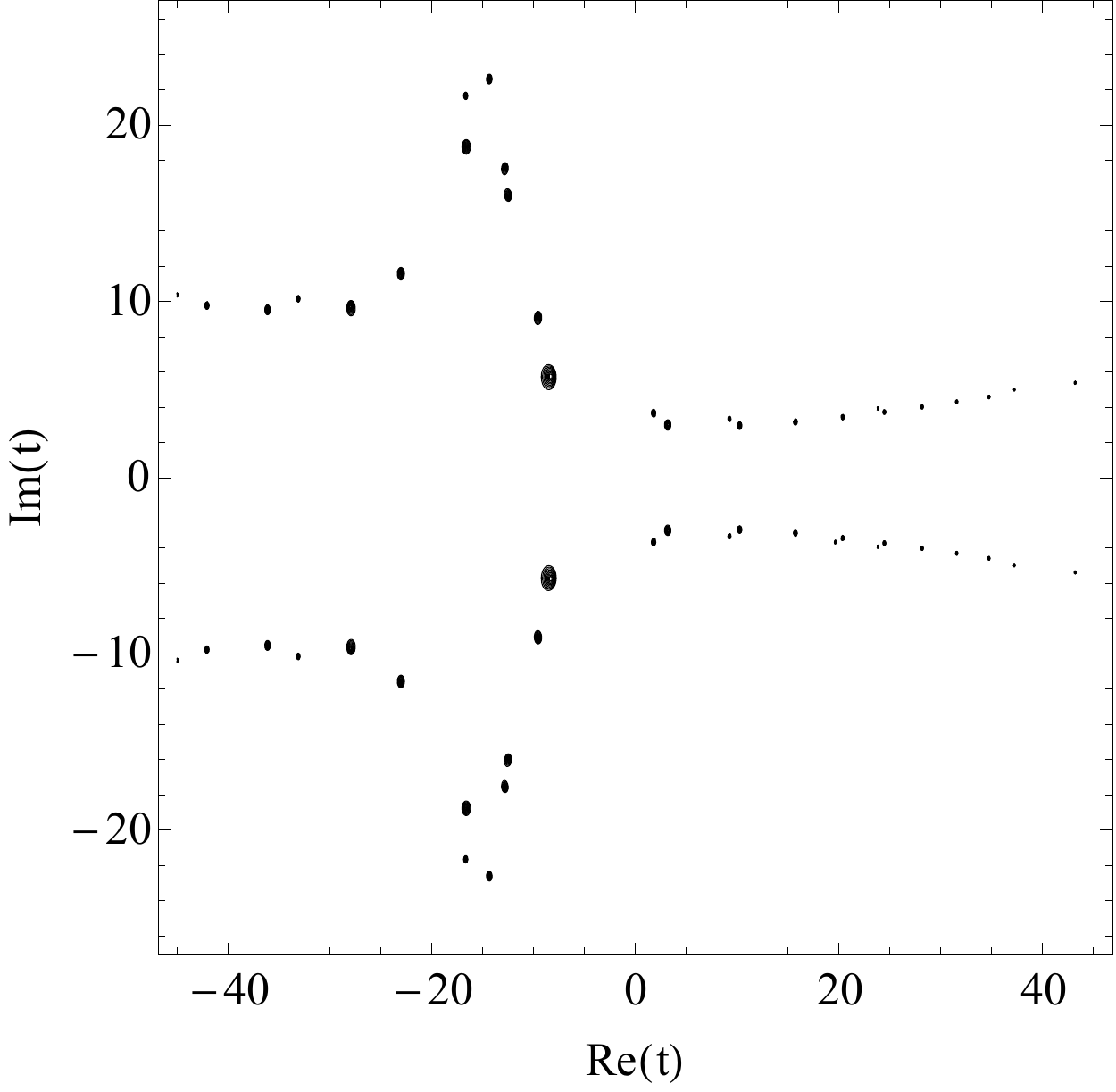}
  \end{minipage}
  \begin{minipage}{7cm}
   \includegraphics[width=7cm]{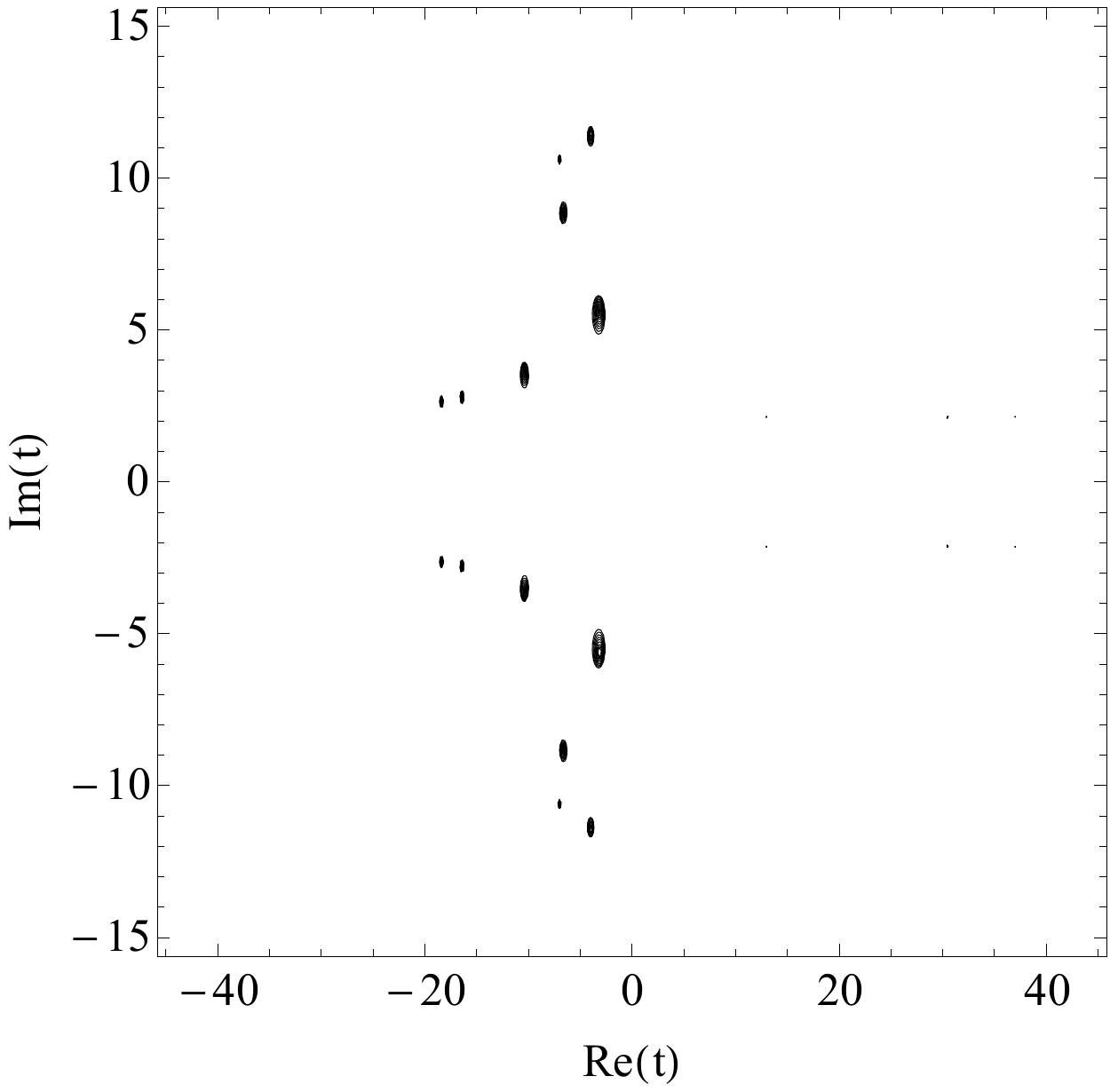}
  \end{minipage}
  \caption{Contour plots of $|\Omega(\mathbf{q},t)|^{2}$ in the complex $t$ plane,
  showing the location of turning points where $\Omega(\mathbf{q},t)=0$.
  These plots are for the elliptic polarization $\delta=0.5$.
  The other field parameters are the same as in Fig.~\ref{fig:1}.
  From top left to bottom right the values of the large chirp parameters are $b=0, 0.005, 0.02, 0.06$
  $m^2$, respectively,
  and the  momentum values (in units of $m$) are
  $(q_{x}=0,q_{y}=0) , (q_{x}=0.1,q_{y}=0.3), (q_{x}=-0.25,q_{y}=-0.5), (q_{x}=0.06,q_{y}=0.06)$,
   respectively.}
  \label{fig:10}
\end{figure}

\begin{figure}
\centering
  \begin{minipage}{7cm}
   \includegraphics[width=7cm]{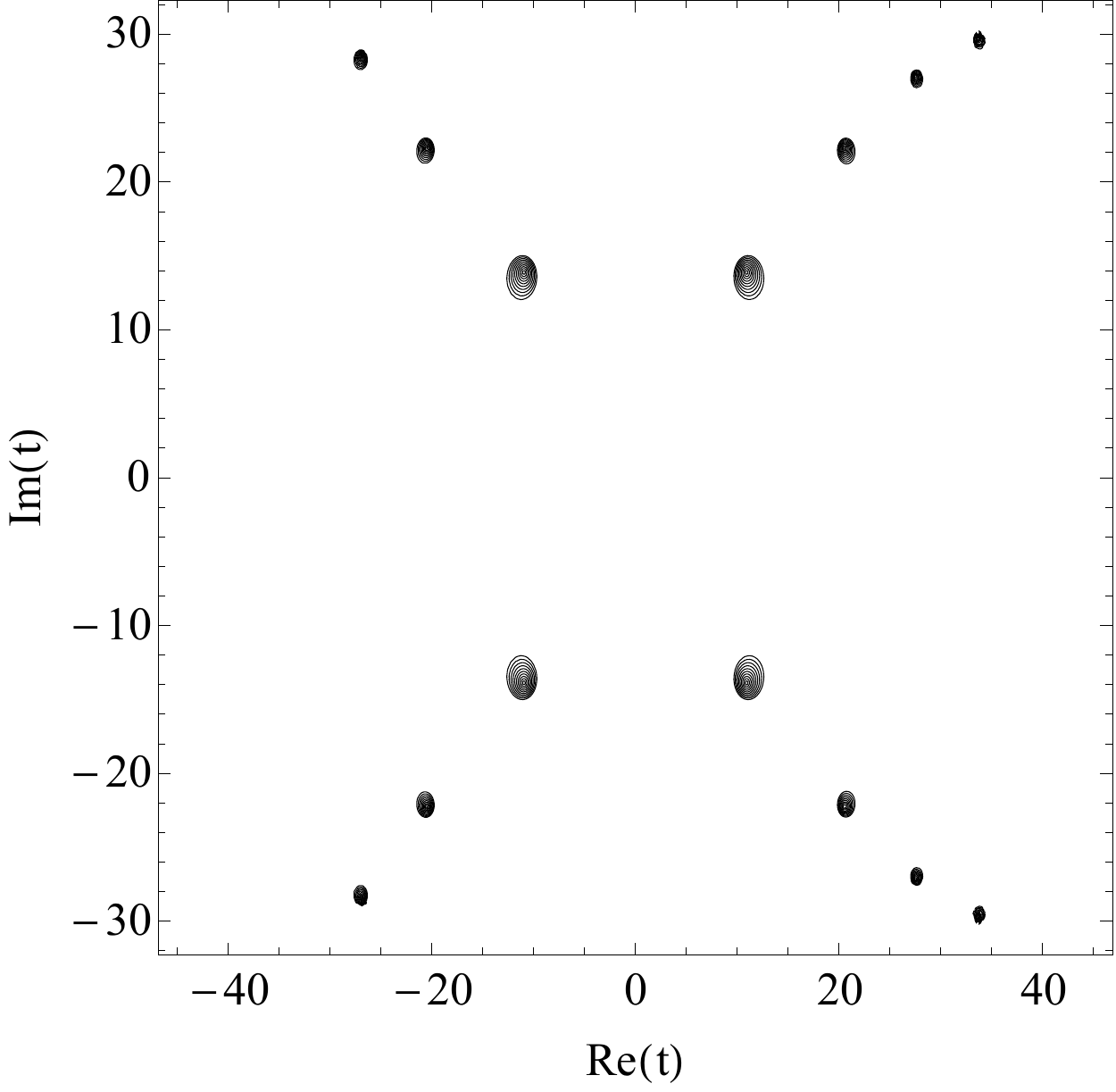}
  \end{minipage}
  \begin{minipage}{7cm}
    \includegraphics[width=7cm]{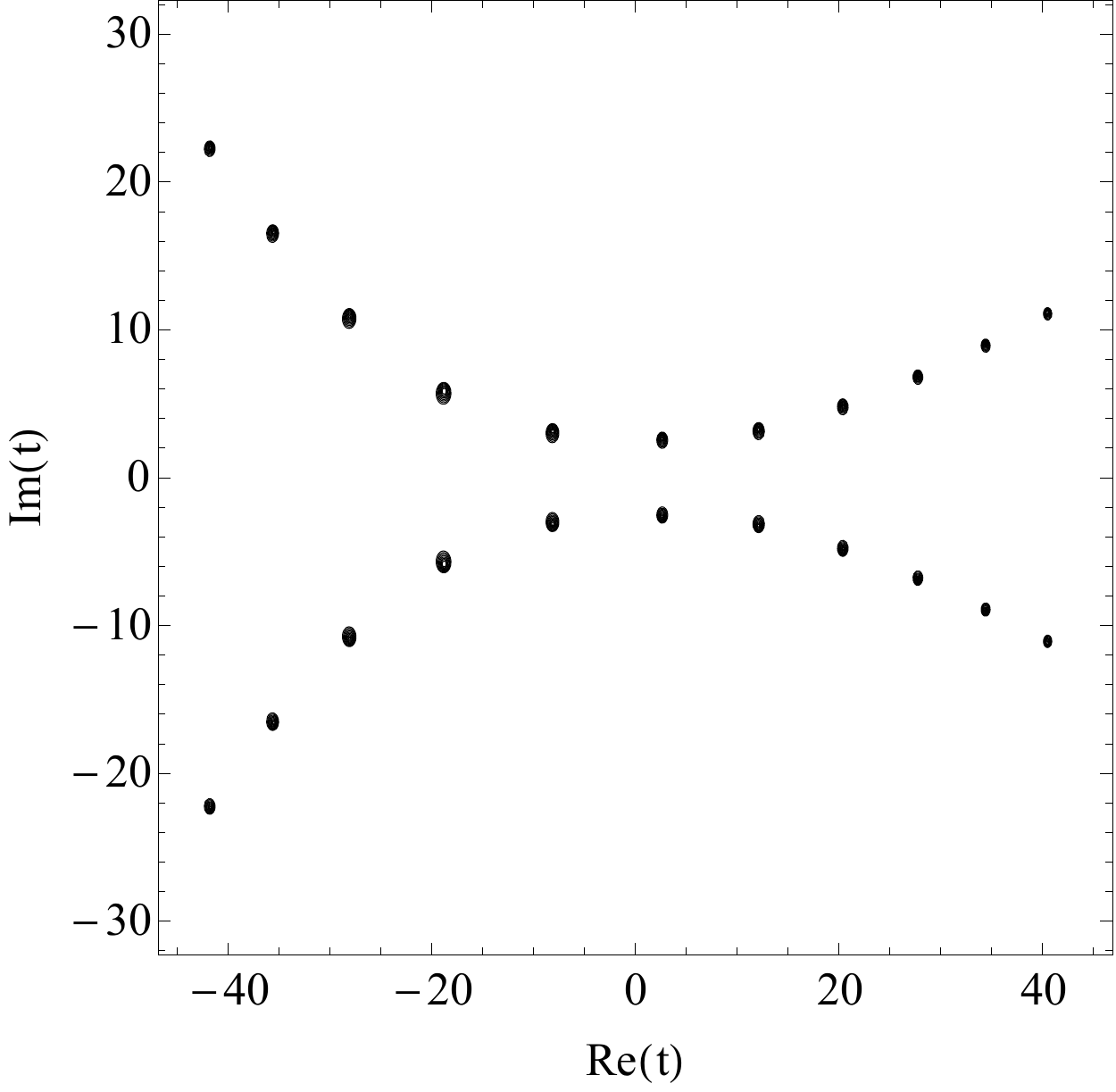}
  \end{minipage}

  \begin{minipage}{7cm}
   \includegraphics[width=7cm]{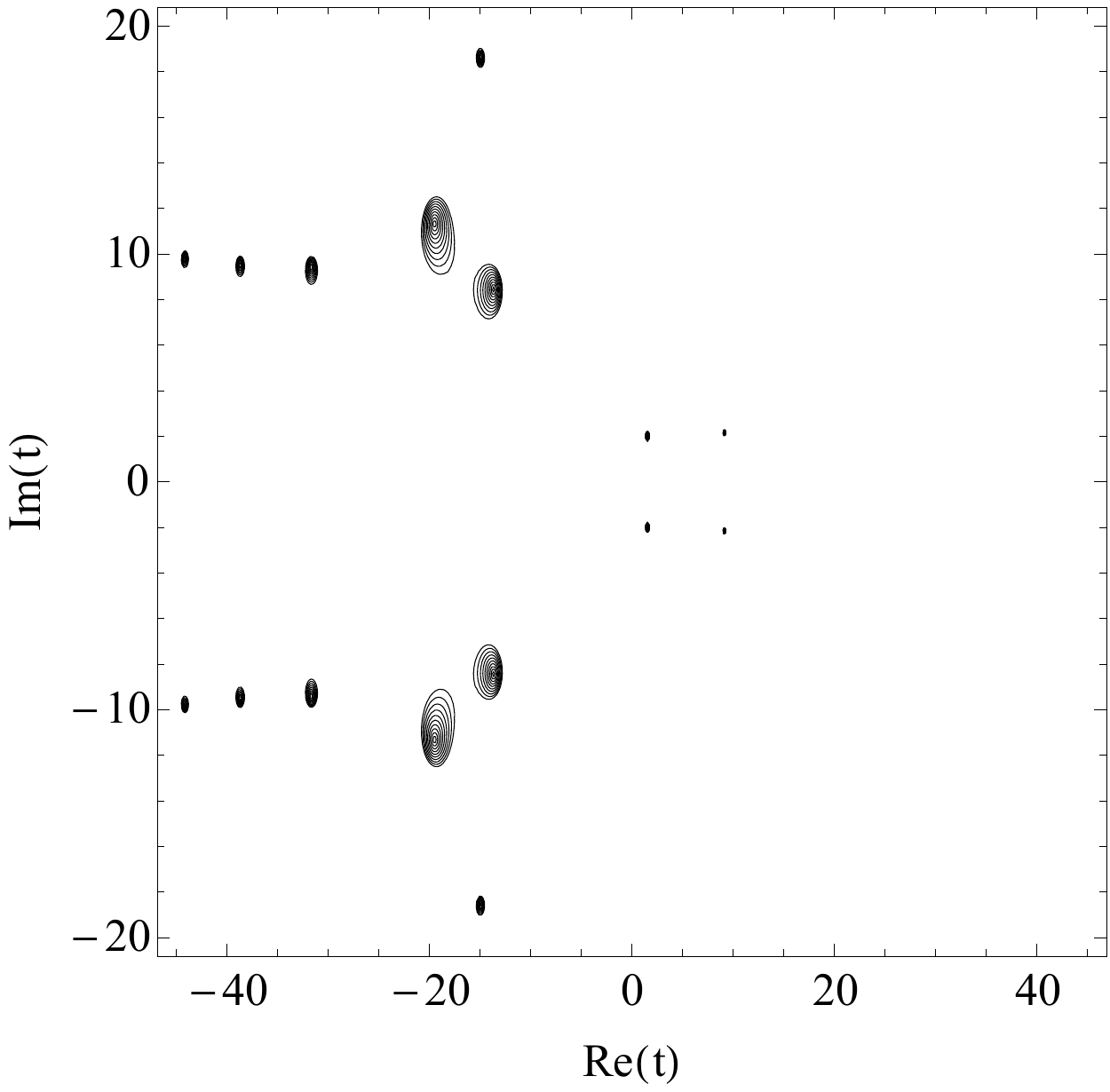}
  \end{minipage}
  \begin{minipage}{7cm}
   \includegraphics[width=7cm]{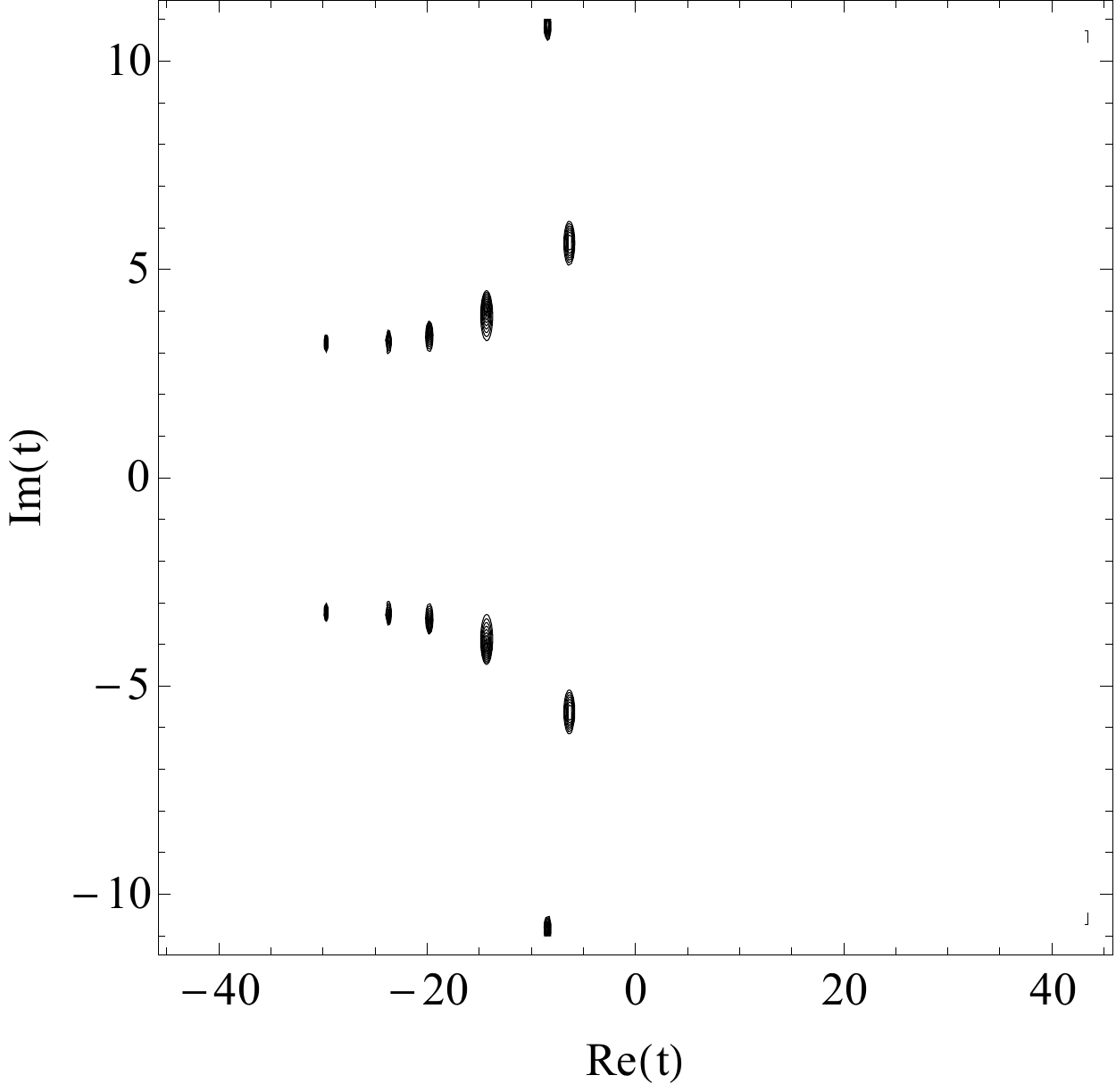}
  \end{minipage}
  \caption{Contour plots of $|\Omega(\mathbf{q},t)|^{2}$ in the complex $t$ plane,
  showing the location of turning points where $\Omega(\mathbf{q},t)=0$.
  These plots are for the circular polarization $\delta=1$.
  The other field parameters are the same as in Fig. \ref{fig:1}.
  From top left to bottom right the values of the large chirp parameters are $b=0, 0.005, 0.03, 0.06$
  $m^2$, respectively,,
  and momentum values (in units of $m$) are
  $(q_{x}=0,q_{y}=0) , (q_{x}=-0.5,q_{y}=0), (q_{x}=0.3 ,q_{y}=0.25), (q_{x}=0.2,q_{y}=0)$, respectively.}
  \label{fig:11}
\end{figure}

For the gauge potential of the electric field given in eq.~\eqref{eq1} there exists an infinite number of
complex turning points which can be obtained only numerically. In addition, as for non-vanishing
chirp parameters the momentum spectra are due to the strong interference effects
peaked at quite different momenta we follow here ref.~\cite{Dumlu:2010vv} and select for each
parameter set representative values for the momenta to evaluate the turning points.

For the case of linear polarization, these turning points are shown in Fig. \ref{fig:9}.
For $b=0$, there is an infinite tower of turning point pairs,
but only the closest ones to the real axis contribute effectively.
(From Fig. \ref{fig:5}(a) one can infer the weak interference patterns which relate
to the exponentially suppressed contributions of the other turning points.)
As the chirp parameters change the corresponding dominating pairs of turning points close
 to the real axis are for small values of $b$ altered mildly but for larger values
 very strongly, see Fig.\ref{fig:9}. Hereby several pairs of turning points possess a
 similar distance to the real axis, and therefore the appearance of strong interference
 effects is understood.
For elliptic polarization practically the same overall picture applies, see  Fig.~\ref{fig:10}
with, however, some differences in turning points positions, especially for large values of
$b$.

For circular polarization the turning points are depicted in Fig. \ref{fig:11}
for several values of chirp parameters. As we can see, for $b=0$, one of the important difference
from the cases of linear polarization and elliptic polarization shown in Fig. \ref{fig:9}(a) is that now
the dominant contribution comes from the two central turning points because
 they are equally distant from the real axis. The two turning point pairs explain
the (weak) interference effect observed for circular polarization and $b=0$.
Increasing now the chirp parameter leads now to quite a distinctive pattern of turning points
as for the other two polarizations, and explains thus the differences in response to the
chirp parameter for circular versus the linear or a more general polarization.

\section{Summary and Conclusions}

Within the real-time DHW formalism we studied the effect of linear frequency chirps for
four polarizations, namely linear, elliptic, near-circular elliptic and circular polarization.
The main results for the number densities and spectra of produced pair
can be summarized as follows:

For an electric field with linear polarization field the produced pairs' spectra exhibit
a shift and split of peaks as well as strong interference effects as the  chirp
parameter increases. The most complex pattern for increasing chirps occurs,
not unexpectedly, for an elliptic polarization. For the near-circular elliptic and
the circular polarization the ring form of the spectrum present at vanishing chirps is
distorted, spiral structures appear and eventually, for very large chirp parameters,
the peak is shifted to the central region.

The most important finding, however, is the very strong increase in number densities
when the chirp parameter is increased. For vanishing and small chirps we have verified
the known differences in number densities for different polarizations, with the largest
number density achieved by a linearly polarized field. Also quite unexpected, this effect
goes away for larger chirps, and the number densities for different polarizations become
degenerate.

In this exploratory study we restricted ourselves to a quite large value of the electric
field and a quite short pulse duration. To verify or falsify
a possible interpretation of the increase in the number of produced
pairs in terms of an onset of multi-photon pair production, a study with much longer pulses
would be necessary. Given the steep increase in the produced number pair densities
and the related improved potential for an experimental observation the effort of
a study employing smaller values of the electric field and much longer pulse times
 is certainly  justified.

\newpage

\begin{acknowledgments}

\noindent
We thank Christian Kohlf\"urst for a critical reading of the manuscript and helpful
remarks.

\noindent
This work was supported by the National Natural Science Foundation of China (NSFC) under
Grant No.\ 11475026 and 11875007.
The computation was carried out at the HSCC of the Beijing Normal University.

\end{acknowledgments}

\end{document}